\documentclass[lettersize,journal]{IEEEtran}
\usepackage{amsmath,amsfonts,amssymb,stmaryrd}
\usepackage{array}
\usepackage[caption=false,font=footnotesize,labelfont=sf,textfont=sf]{subfig}
\usepackage{textcomp}
\usepackage{stfloats}
\usepackage{url}
\usepackage{verbatim}
\usepackage{graphicx}
\usepackage{authblk}
\usepackage{stackengine}
\usepackage{balance}

\usepackage[table,xcdraw]{xcolor}
\usepackage{color,soul}
\usepackage{threeparttable}
\usepackage{xcolor}
\usepackage[hidelinks]{hyperref}
\usepackage{adjustbox}
\usepackage{float}
\definecolor{Paired-1}{RGB}{31,120,180}
\definecolor{Paired-2}{RGB}{166,206,227}
\definecolor{Paired-3}{RGB}{51,160,44}
\definecolor{Paired-4}{RGB}{178,223,138}
\definecolor{Paired-5}{RGB}{227,26,28}
\definecolor{Paired-6}{RGB}{251,154,153}
\definecolor{Paired-7}{RGB}{255,127,0}
\definecolor{Paired-8}{RGB}{253,191,111}
\definecolor{Paired-9}{RGB}{106,61,154}
\definecolor{Paired-10}{RGB}{202,178,214}
\definecolor{Paired-11}{RGB}{177,89,40}
\definecolor{Paired-12}{RGB}{255,255,153}
\definecolor{dgreen}{RGB}{3,125,60}
\definecolor{dblue}{RGB}{47,82,143}
\definecolor{bgreen}{RGB}{0,128,128}
\definecolor{bgred}{RGB}{237,29,33}
\usepackage{bm}
\usepackage{dsfont}
\usepackage{numprint}

\usepackage{adjustbox}
\usepackage{booktabs}
\usepackage{multirow}

\usepackage{cite}
\usepackage{algpseudocode}
\usepackage{caption}
\usepackage{balance}
\usepackage[utf8]{inputenc}

\usepackage{graphicx}
\usepackage{textcomp}
\usepackage{pgfplots, pgfplotstable}
\usepackage{tikz}
\usetikzlibrary{spy} 
\usetikzlibrary{matrix, positioning, patterns, shapes, arrows}
\usetikzlibrary{positioning}
\usepackage{tikzscale}
\usepgfplotslibrary{statistics}
\usepackage{bchart}
\pgfplotsset{compat=newest}
\usepgfplotslibrary{groupplots}
\usepgfplotslibrary{colorbrewer}
\usepackage[linesnumbered,ruled,vlined]{algorithm2e}

\SetCommentSty{mycommfont}
\usepackage{lipsum}

\usepackage{microtype}

\usepackage{etoolbox}
\patchcmd{\thebibliography}{\leftmargin\labelwidth}
    {\itemsep -1.4pt \leftmargin\labelwidth}{}{}{}

\pgfplotsset{every x tick label/.append style={font=\scriptsize}}
\pgfplotsset{every y tick label/.append style={font=\scriptsize}}

\makeatletter
\newcommand\thefontsize[1]{{#1 The current font size is: \f@size pt\par}}
\makeatother

\makeatletter
\def\bstctlcite{\@ifnextchar[{\@bstctlcite}{\@bstctlcite[@auxout]}}
\def\@bstctlcite[#1]#2{\@bsphack
  \@for\@citeb:=#2\do{%
    \edef\@citeb{\expandafter\@firstofone\@citeb}%
    \if@filesw\immediate\write\csname #1\endcsname{\string\citation{\@citeb}}\fi}%
  \@esphack}
\makeatother 

\newcommand\scalemath[2]{\scalebox{#1}{\mbox{\ensuremath{\displaystyle #2}}}}

\graphicspath{{figs/}}

\begin{document}
\title{Hardware Implementation of Projection-Aggregation Decoders for Reed-Muller Codes}



\author{\IEEEauthorblockN{Marzieh Hashemipour-Nazari, Andrea Nardi-Dei, Kees Goossens, and Alexios Balatsoukas-Stimming}\\
              \IEEEauthorblockA{Electronic Systems, Eindhoven University of Technology, The Netherlands\\}
              }

\maketitle

\begin{abstract}
This paper presents the hardware implementation of two variants of projection-aggregation-based decoding of Reed-Muller (RM) codes, namely unique projection aggregation (UPA) and collapsed projection aggregation (CPA). 
Our study focuses on introducing hardware architectures for both UPA and CPA. Through thorough analysis and experimentation, we observe that the hardware implementation of UPA exhibits superior resource usage and reduced energy consumption compared to CPA for the vanilla IPA decoder. This finding underscores a critical insight: software optimizations, in isolation, may not necessarily translate into hardware cost effectiveness. 
\end{abstract}


\section{Introduction}
\label{sec:intro}

\IEEEPARstart{T}{he} demand for ultra-reliable low-latency communications (URLLC) in future communication systems, particularly for machine-type communications (MTC) and IoT scenarios, necessitates the development of efficient error-correction coding schemes capable of handling short packets \cite{Sisi2024,Durisi2016,Mahmood2020}. 
Traditional codes like low-density parity check (LDPC) and turbo codes can face challenges in maintaining high reliability for short packets due to their design complexities and  inherent limitations\cite{gallager1963low,berrou1996near}.
Polar codes \cite{Arikan2009} offer a promising solution for short blocks through successive cancellation list (SCL) decoding~\cite{Tal2015}. However, optimizing their performance for short block lengths necessitates adjustments such as using a large list size, which increases decoding complexity, and the integration of cyclic redundancy check (CRC) codes, which decreases the effective information rates, especially for low-rate codes.

Reed-Muller (RM) codes, initially introduced by Reed~\cite{Reed1954} and Muller~\cite{Muller1954}, have regained attention due to the practical limitations of polar codes, particularly for short block lengths \cite{Arikan2008,Arikan2010}. 
Unlike polar codes, which aim to minimize error probability under successive cancellation (SC) decoding \cite{Tal2013,Trifonov2012}, RM codes are designed to optimize performance under maximum likelihood (ML) decoding by maximizing the minimum distance between codewords. Consequently, RM codes exhibit superior performance under ML decoding compared to polar codes. Recent research has also highlighted  capacity-achieving capability of RM codes for binary erasure channels (BEC) \cite{Kudekar2017}, binary symmetric channels (BSC) \cite{Abbe2015}, and general binary-input memoryless symmetric (BMS) channels \cite{Reeves2024}.

Several decoding algorithms have been developed to efficiently decode RM codes \cite{Sidel1992,Sakkour2005,Dumer2004,Dumer2006,Santi2018, Ivanov2019}. Moreover, due to the structural similarities between RM codes and polar codes, improved decoding algorithms initially designed for polar codes, such as fast successive cancellation (FSC) and fast successive cancellation list (FSCL) decoding, can be adapted to RM codes \cite{Hashemi2018,Ardakani2019}. Recently, recursive projection-aggregation (RPA), a near-maximum likelihood (ML) decoding technique, has demonstrated significant potential to improve error-correction performance for short block-length RM codes \cite{Ye2020}. 
The RPA decoding method first breaks down the received codeword into shorter RM codewords using codeword projections.
Subsequently, it recursively decodes these projected codewords until reaching first-order codewords, which are efficiently decoded using the fast Hadamard transform (FHT) \cite{Beext86}. Finally, it aggregates the shorter decoded codewords to reconstruct an estimate of the original received codeword.

The computational complexity of RPA decoding depends on the number of projections required to generate first-order codewords, which is influenced by the order of the RM code $r$ and its block length $n$. 
The complexity of RPA decoding scales as $n^r$, making it generally impractical for RM codes with $r>2$. Furthermore, the recursive nature of RPA poses challenges for hardware implementation. Consequently, various approaches have been proposed to address these issues.

The work of \cite{JiaJie2021} introduces a method to reduce complexity by incorporating an early stopping condition. However, this approach necessitates additional hardware to verify the early stopping condition.
The work of\cite{Fathollahi2021} proposes $K$-sparse RPA (K-SRPA), employing multiple pruned RPA decoders instead of one full RPA decoder and utilizing a CRC code to select the best output among the outputs of $k$ decoders. As explained previously, integrating a CRC into the codeword decreases the information rate, and the presence of multiple decoders increases hardware cost. 
The work of \cite{Voigt2023} suggests reducing the complexity of \cite{Fathollahi2021} by incorporating run-time preprocessing computations on the received vector to select the optimal projections for pruned RPA, thus reducing the number of projections compared to K-SRPA.
However, from a hardware implementation perspective, this method would require extensive run-time preprocessing components, increasing latency and hardware resources while also disrupting the inherent parallelism of the RPA decoder.
The research conducted in \cite{Lian2020} introduces collapsed projection aggregation (CPA), merging $r-1$ levels of projections into one level, resulting in fewer first-order codewords compared to RPA decoding.
The work of \cite{Hashemipour2022} combines the concepts of \cite{JiaJie2021} and \cite{Fathollahi2021} and presents a multi-factor pruned RPA (MFP-RPA), which achieves good performance while reducing complexity. However, its unstructured pruning method poses challenges for hardware implementation.
Subsequently, based on observations in MFP-RPA, \cite{Hashemipour2023} proposes a systematic pruning approach called recursive unique projection-aggregation (RUPA). RUPA prevents the generation of duplicate projections in RPA, resulting in exactly the same algorithmic complexity as CPA.
In addition, pruned CPA (PCPA)~\cite{Huang2022} and optimized PCPA~\cite{Jiajie2022} proposed methods to reduce CPA complexity. However, similar to \cite{Voigt2023}, these methods require run-time preprocessing to select the best group of projections based on the received vector.

In addition to the considerable computational complexity of the RPA algorithm, its recursive nature poses significant challenges for hardware implementation. To address this, \cite{Hashemipour-Nazari2021} introduces the iterative projection aggregation (IPA) algorithm, converting the recursive structure of the RPA decoder into an iterative one. Recently, in \cite{HashemipourTCAS2023}, a soft-input flexible pipelined architecture for the IPA algorithm has been proposed. Post-synthesis results in \cite{HashemipourTCAS2023} indicate that using the proposed IPA architecture, RM codes outperform polar codes under a state-of-the-art SCL decoder~\cite{Ren2022} with identical block length, information rate, and error-correction performance. However, implementing the IPA decoder for RM codes with $r>2$ remains challenging due to either high resource usage or high latency in parallel or sequential
configurations, respectively. 
To extend the applicability of IPA to a broader range of RM codes, algorithmic complexity-reduction methods are needed. 

\subsubsection*{Contributions \& Outline} In this work, we first highlight the challenge posed by the unique projection selection method~\cite{HashemipourTCAS2023} for hardware implementation and then address it by formulating and solving an integer linear problem (ILP) for projection selection. We then propose a flexible hardware architecture tailored to the iterative unique projection aggregation (IUPA) decoder, leveraging the ILP solution to minimize resource usage. 
Additionally, we introduce a semi-flexible pipelined architecture for CPA decoding~\cite{Lian2020}, facilitating a comprehensive analysis and comparison of the IUPA, CPA, and IPA architectures. Our results are a reminder that focusing solely on improving the algorithmic complexity might not always result in better hardware. For instance, although CPA decoding has lower asymptotic computational complexity than IPA, it may still require more hardware resources in certain configurations.

The remainder of this paper is organized as follows. In Section~\ref{backgnd}, we provide an overview of RM codes and the decoding algorithms associated with them, including RPA, IPA, CPA, and IUPA. We then highlight the challenges of implementing the IUPA algorithm in hardware and propose solutions in Section~\ref{sec:IUPA_imp}, while in Section~\ref{sec:iupa:arch} we introduce a flexible hardware architecture tailored to IUPA. Following this, we present the hardware architecture for CPA decoding in Section~\ref{sec:CPA_imp}.
In Section~\ref{sec:result}, we present comprehensive simulation and synthesis results, offering a thorough comparison of IPA, IUPA, and CPA decoders. Finally, Section~\ref{sec:conclusion} concludes the paper and summarizes our findings.

\section{Background}
\label{backgnd}
\subsubsection*{Notation} Throughout this paper, vectors are represented in boldface. Matrices are denoted by bold uppercase non-italic letters. The notation $y(i)$ refers to the $i$-th element of the vector $\boldsymbol{y}$. Additionally, $D_{ij}$ denotes the element on row $i$ and column $j$ of the matrix $\mathbf{D}$.
$\mathbf{D}_{i\star}$ and $\mathbf{D}_{j\star}$ indicate the $i$-th row and the $j$-th column in matrix $\mathbf{D}$, respectively. 
Furthermore,  the symbol $\oplus$ signifies summation over the binary field $\mathbb{F}_2$.
Finally, ${\binom{m}{s}}_2$ is the $2$-binomial coefficient and is given by:
\begin{equation}
{\binom{m}{s}}_2 =\prod_{i=0}^{s-1} \frac{1-2^{m-i}}{1-2^{i+1}}.
\end{equation}
\subsection{Reed-Muller codes}
Reed-Muller (RM) codes denoted as RM$(m,r)$ are linear block codes, where $r$ denotes the code rate and $m$ determines the block length with $n = 2^m$.
They  encode a $k$-bit binary vector $\boldsymbol{u}$ into an $n$-bit codeword $\boldsymbol{c}= \boldsymbol{u}\mathbf{G}_{(m, r)}$, where  $k=\sum_{i=0}^r \binom{m}{i}$ and $\mathbf{G}_{(m, r)}$ is the generator matrix of the RM$(m,r)$ code.
The matrix $\mathbf{G}_{(m, r)}$ can be constructed recursively as~\cite{Plotkin1960}:
\begin{equation}\label{eq:GMtxRM}
\mathbf{G}_{(m, r)}= \begin{bmatrix}
  \mathbf{G}_{(m-1, r)} & \mathbf{G}_{(m-1, r)}\\ 
  \mathbf{0} & \mathbf{G}_{(m-1, r-1)}
\end{bmatrix},
\quad
\mathbf{G}_{(1, 1)}{=} \begin{bmatrix}
  1 & 1\\ 
  0 & 1
\end{bmatrix}.
\end{equation}
Moreover, each codeword $\boldsymbol{c}=(c(\boldsymbol{z}),\boldsymbol{z}\in \mathbb{E})\in \text{RM}(m,r)$ represents a binary incidence vector of a monomial $F$ with a degree at most $r$, evaluated across all elements $z_i$, where $i \in {0,1,\dotsc,2^m-1}$, of $\mathbb{E}:= \mathbb{F}^{m}_{2}$.

\subsection{RPA decoding}
\label{sec:rpa}

The work of \cite{Ye2020} demonstrates that projecting $\boldsymbol{c}\in \text{RM}(m,r)$ onto the cosets of a one-dimensional subspace $\mathbb{B}$ of $\mathbb{E}$ generates the codeword $\boldsymbol{c}_{/ \mathbb{B}}\in \text{RM}(m-1,r-1)$. The projection of $\boldsymbol{c}$ onto the cosets of $\mathbb{B}$ is defined as follows:
\begin{equation}
\label{eq:proj}
\boldsymbol{c}_{/ \mathbb{B}}=\operatorname{Proj}(\boldsymbol{c}, \mathbb{B}):=\left(c_{/ \mathbb{B}}(T):= {\oplus}_{z\in T}c(z), T  \in \mathbb{E} / \mathbb{B} \right),
\end{equation} 
where  $\mathbb{E} / \mathbb{B}$ is a quotient space comprising $2^{m-1}$ cosets $T$ of $\mathbb{B}$. The construction of $\mathbb{E} / \mathbb{B}$ is detailed in~\cite{Ye2020}.
RPA decoding leverages this characteristic of RM codes and has three main steps: projection, recursive decoding, and aggregation. 
\subsubsection{Projection}
Let $\boldsymbol{L}$ denote the vector of log-likelihood ratios (LLRs) corresponding to the output of a transmitted codeword $\boldsymbol{c} \in \text{RM}(m,r)$. RPA decoding computes the projected vector $\boldsymbol{L}_{/ \mathbb{B}}$ for each one-dimensional subspace $\mathbb{B}$ of $\mathbb{E}:= \mathbb{F}^{m}_{2}$. This projection can be computed using the hardware-friendly min-sum (MS) approximation~\cite{JiaJie2021}:
\begin{equation}
\label{eq:minsum}
\scalemath{0.85}{
\boldsymbol{L}_{/ \mathbb{B}} \approx \left(c_{/ \mathbb{B}}([T]):=\min_{z\in T}\left\lbrace|L(z)|\right\rbrace \prod_{z\in T} \text{sign}\left(L\left(z\right)\right), T  \in \mathbb{E} / \mathbb{B} \right).
}
\end{equation}
Since each one-dimensional subspace of $\mathbb{E}$ is comprised of $0$ and a non-zero element, there exist $2^m-1$ such subspaces in total, leading to $2^m-1$ projected vectors.

\subsubsection{Recursive decoding}
RPA calls itself recursively until $r = 1$, i.e., until first-order codewords are obtained. Subsequently, it employs an efficient fast Hadamard transform (FHT) decoder~\cite{Green66} to decode the generated first-order RM codes.
\subsubsection{Aggregation}
After the projection and recursive decoding steps, $2^m-1$ LLR vectors $\hat{\boldsymbol{L}}_{/ \mathbb{B}}$ are obtained. RPA aggregates these codewords with the input vector $\boldsymbol{L}$ as:
\begin{equation}\label{eq:agg}
{\hat{L}}(z)=\frac{1}{2^m-1}\sum_{i=1}^{2^m-1}\left(1-2\hat{c}_{/ \mathbb{B}_i}(\left[z \oplus\mathbb{B}_i\right])\right)L(z\oplus i),
\end{equation}
where $\hat{L}(z)$ denotes the $z$-th coordinate of ${\boldsymbol{\hat{L}}}=\left({\hat{L}}(z),z\in \mathbb{E}\right)$, and $\left[z \oplus\mathbb{B}_i\right]$ is the index of the coset $\mathbb{B}_i$ that includes coordinate $z$. Additionally, $\boldsymbol{\hat{c}}_{/ \mathbb{B}_i}$ represents the result of recursive decoding of the projected vector on subspace $\mathbb{B}_i$. Then, $\boldsymbol{L}$ is replaced by $\boldsymbol{\hat{L}}$ and additional RPA iterations are performed until $\boldsymbol{\hat{L}}$ is equal to $\boldsymbol{L}$ or a maximum number of iterations is reached. Finally, RPA decoding takes element-wise hard decisions on ${\boldsymbol{\hat{L}}}$ to determine the binary decoded codeword $\hat{\boldsymbol{c}}$.

The number of first-order codewords generated in the innermost level of recursion is a proxy for the computational complexity of the RPA algorithm and scales exponentially with $r$~\cite{Ye2020}. 
From an implementation perspective, this makes the RPA algorithm impractical  for $r>2$.

\subsection{IPA decoding and implementation}
\label{sec:ipa}
The simulation results in \cite{Hashemipour-Nazari2021} show that the majority of projected codewords are corrected after the initial iteration within each recursion level. Moreover, the work of\cite{HashemipourTCAS2023} demonstrates that internal iterations disrupt the intrinsic parallelism of the RPA decoding method, leading to difficulties in hardware implementation.
Therefore, the iterative projection aggregation (IPA) decoding method eliminates internal iterations, keeping only the iterations at the outer level. Based on this approach, both hard-decision and soft-decision IPA decoders were proposed and implemented in~\cite{Hashemipour-Nazari2021} and~\cite{HashemipourTCAS2023}, respectively.

\begin{figure}[t]
\centering
\includegraphics[width=0.5\textwidth]{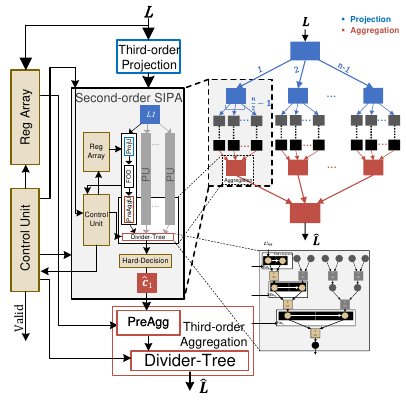}
\caption{Third-order IPA decoder for one iteration~\cite{HashemipourTCAS2023}. }
\label{fig:SIPABD}
\end{figure}
Fig.~\ref{fig:SIPABD} illustrates the block diagram of the (soft-input) IPA decoder for third-order RM codes that we proposed in \cite{HashemipourTCAS2023}. Specifically, the architecture proposed in \cite{HashemipourTCAS2023} employs a foundational decoder for second-order codes, which is comprised of $p$ processing units (PUs), each containing a projection unit (\textit{ProjU}), a first-order decoder (\textit{FOD}), and a pre-aggregation unit (\textit{PreAggU}).
The second-order IPA decoder can be configured from fully sequential with $p=1$ PU to fully parallel with $p=2^m-1$ PUs.
The \textit{ProjU} component for the $i$-th PU encompasses $\frac{2^m}{p}$ distinct permutations based on corresponding subspaces $\mathbb{B}_j, j\in \{i,\dotsc,i\times \frac{2^m}{p}-1\}$, for the received vector $\boldsymbol{L}\in\text{RM}(m,2)$. Additionally, it includes $2^m-1$ two-input \textit{MinSum} components, implementing the MS update rule of~\eqref{eq:minsum} on each pair of permuted coordinates of $\boldsymbol{L}$. The \textit{FOD} component is a hardware implementation of the FHT-based first-order decoder.

The aggregation step based on \eqref{eq:agg} requires all decoded results of the first-order codewords for all $2^m-1$ projections to compute an average for each coordinate of $\boldsymbol{\hat{L}}$. Due to this dependency, it cannot be implemented directly in parallel. Hence, \cite{HashemipourTCAS2023} divided it into two steps: pre-aggregation and averaging. 
In the pre-aggregation step, $\boldsymbol{L}^{i}_{\text{agg}}$ for the decoded projected vector $\hat{\boldsymbol{c}}_{/\mathbb{B}_i}$ is calculated as:
\begin{equation}\label{eq:preagg}
{{L}^i_{\text{agg}}}(z):=\left(1-2\hat{c}_{/ \mathbb{B}_i}(\left[z \oplus\mathbb{B}_i\right])\right)L(z\oplus i).
\end{equation}
The \textit{PreAggU} component includes two crossbars for selecting the proper indices for $\hat{c}_{/ \mathbb{B}_i}$ and $\boldsymbol{L}$.
Moreover, since $\left(1-2\hat{c}_{/ \mathbb{B}_i}(\left[z \oplus\mathbb{B}_i\right])\right)$ is either $1$ or $-1$ the product in \eqref{eq:preagg} can be performed with a two's complement circuit for $L(z\oplus i)$  with $\hat{c}_{/ \mathbb{B}_i}\left(\left[z \oplus\mathbb{B}_i\right]\right)$ as its enable port.
To finalize the aggregation process and compute the average of all estimated vectors $\boldsymbol{L}^{i}_{\text{agg}}, i\in\{1,\dotsc,n-1\}$, a divider tree composed of $m$ two-input adders and shift registers is employed. Further details on the divider tree are provided in \cite[Section~III-B]{HashemipourTCAS2023}.

Post-synthesis analysis of IPA has shown that it outperforms a state-of-the-art SCL decoder~\cite{Ren2022} for polar codes of identical block-length and information rate, achieving comparable error-correction performance for selected short codes.
However, its high number of projections either requires large hardware for RM codes with $r>2$ in low-latency configurations or has high latency in low-resource setups.

\subsection{CPA decoding}
\label{Sec:CPA}
Collapsed projection aggregation (CPA)~\cite{Lian2020} decoding directly projects the received vector of LLRs $\boldsymbol{L}$ onto $(r-1)$-dimensional subspaces of $\mathbb{E}$ instead of  $r-1$ levels of one-dimensional projections.
In CPA decoding, the projection rule is similar to \eqref{eq:minsum}. However, each coset $T$ in CPA contains $2^{r-1}$ elements, unlike the IPA algorithm where each coset contains only two elements.
Since, there exist $\binom{m}{r-1}_2$ different $(r-1)$-dimensional subspaces for $\mathbb{E}$, CPA decoding yields $\binom{m}{r-1}_2$ distinct first-order codewords instead of $\prod_{i=0}^{r-2}{\binom{m-i}{1}_2}$ in the IPA decoding.
After decoding the obtained first-order codewords using an FHT-based decoder, CPA constructs the vector ${\boldsymbol{\hat{L}}}=\left({\hat{L}}(z),z\in \mathbb{E}\right)$ according to:
\begin{equation}\label{eq:agg:cpa}
\scalemath{0.85}{
{\hat{L}}(z) = \frac{1}{n_p} \sum_{i=1}^{n_P} {-1^{\hat{c}_{/ \mathbb{B}_i}([T])} 
\left(\min_{z_{j}\in T-z}\lbrace|L(z_j)|\rbrace \prod_{z_{j}\in T-z} \text{sign}\left(L\left(z_{j}\right)\right)\right)}
},
\end{equation}
where $n_P=\binom{m}{r-1}_2$. 
Finally, CPA iterates until either reaching a maximum number of iterations $N_{\text{max}}$ or achieving convergence of the output vector $\boldsymbol{\hat{L}}$ to the input vector $\boldsymbol{L}$.

The complexity of CPA is significantly reduced for RM codes with $r>2$ compared to IPA. This reduction can be quantified using the ratio of the number of first-order codewords generated in CPA ($N_{\text{CPA}}$) to that in IPA ($N_{\text{IPA}}$):
\begin{equation}
\label{NFOD}
\frac{N_{\text{CPA}}}{N_{\text{IPA}}} = \frac{\binom{m}{r-1}_2}{\prod_{i=0}^{r-2}{\binom{m-i}{1}_2}} = \prod_{i=0}^{r-2}{ \frac{1}{2^{i+1}-1}}.
\end{equation}
However, the projection and aggregation rules in CPA are more complex than IPA since they involve $2^{r-1}$ inputs.

\subsection{IUPA decoding}
\label{sec:iupa}
\begin{figure}[t]
\centering
\includegraphics[width=0.4\textwidth]{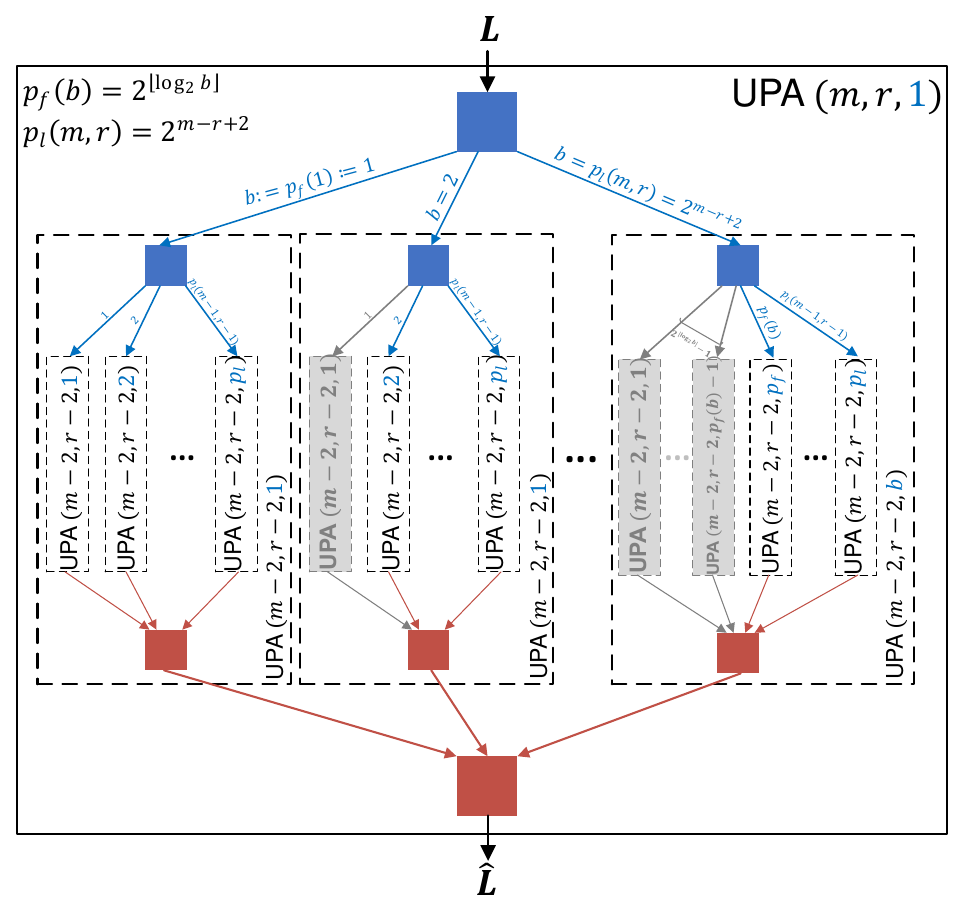}
\caption{Block diagram of two unfolded levels of unique projection aggregation method for RM$(m,r)$~\cite{Hashemipour2023}.}
\label{fig:UPA}
\end{figure}
The work of \cite{Hashemipour2023} reveals that a considerable number of first-order codewords generated after $r-1$ levels of projections in RPA or IPA algorithms are duplicates.
Specifically, within the inner level of projections of both RPA and IPA algorithms, only $\binom{m}{r-1}_2$ unique projected vectors exist. These unique projected vectors are identical to those generated in the CPA algorithm.
By following the RPA projection methodology with $r-1$ levels on one-dimensional subspaces, $N_{\text{IPA}}$ subspaces with $r-1$ dimensions are constructed, of which $\binom{m}{r-1}_2$ are distinct.
Therefore, the unique projection selection method of \cite{Hashemipour2023} proposes a method of projection in $r-1$ levels on one-dimensional subspaces such that there are no duplicate first-order codewords at the innermost level of projection.

The decoding method based on this projection selection is referred to as recursive/iterative unique projection aggregation (RUPA/IUPA). The difference between the recursive and iterative versions of this decoding method lies in the iteration on the levels of recursion in RUPA. As mentioned in Section~\ref{sec:ipa}, internal iterations add complexity to hardware implementation without considerable enhancements in error correction performance~\cite{Hashemipour-Nazari2021}. Therefore, considering the hardware implementation goal of this work, we focus on IUPA.

Fig.~\ref{fig:UPA} illustrates a single iteration of IUPA decoding. The grayed-out branches indicate the redundant projections found in IPA decoding but disregarded in IUPA decoding. 
As depicted in Fig.~\ref{fig:UPA}, the unique projection selection process introduces a new parameter, denoted as $b$, indicating the index of the projection branch. The initial and final projections within each UPA block are labeled as $b_f$ and $b_l$ respectively, and are  defined as:
\begin{equation}
\label{eq:pfl}
\begin{aligned}
& b_f =2^{\left\lfloor\log _2 b\right\rfloor}, \\
& b_l=2^{m-r+2}-1.
\end{aligned}
\end{equation}
IUPA for RM$(m,r)$ starts with $b = 1$ initially and generates $L^i,i\in {1,\hdots,2^{\left\lfloor\log _2 b\right\rfloor}},$ vectors potentially belonging to RM$(m-1,r-1)$ code.
Consequently, IUPA runs on each  $L^i$ with a parameter $b_i = i$.  The number of projections for decoding each $L^i$ is a function of $b_i$, $m$, and $r$, and is defined as $b_l-b_f+1$ according to \eqref{eq:pfl}.
Therefore, the projection and aggregation rules in IUPA are as simple as in IPA and, at the same time, the number of generated first-order codewords in IUPA is identical to that of CPA.

\section{IUPA: Projection Allocation}
\label{sec:IUPA_imp}
IPA decoding avoids duplicate first-order codewords for second-order RM codes, making IUPA decoding identical to IPA.
For higher-order codes, applying the unique selection method to our IPA architecture~\cite{HashemipourTCAS2023} is a promising way to reduce resource usage and energy consumption.
However, implementing the unique selection method in hardware is not straightforward.
In this section, we explain the challenges associated with using the unique selection projection method in hardware for RM$(m,3)$ codes and we then propose a solution.
\subsection{IUPA hardware challenges}
\label{sec:iupa:prob}
Unique projection selection significantly reduces the hardware requirements of the IPA decoder by a factor of $\prod_{i=0}^{r-2}{ \frac{1}{2^{i+1}-1}}$ for a fully-parallel implementation. However, a fully-parallel decoder requires substantial hardware resources and has a high energy consumption, making it impractical~\cite{HashemipourTCAS2023}, so partially-parallel decoders are typically used in practice.

As elaborated in Section~\ref{sec:ipa}, in a partially-parallel implementation of the IPA decoder for RM$(m,3)$ codes, a base second-order decoder with $p$ PUs is defined. This decoder is then reused to decode all the second-order projected vectors, as they all need to project onto the same $2^{m-1}-1$ one-dimensional subspaces. 
The projection and aggregation circuits within each PU of the second-order decoder provide $\tfrac{2^{m-1}}{p}$ distinct permutations of their input vector. 
However, in IUPA decoding for RM$(m,3)$, different second-order codewords require different numbers of projections based on different one-dimensional subspaces.
This non-uniform distribution of projections is not ideal for hardware implementation.
Implementing a second-order decoder providing all possible projections is inefficient in terms of hardware, as only the first projected second-order vector requires all $2^{m-1}-1$ projections, while following vectors require only a subset of projections.
Moreover, as each projection is required at least once, we cannot eliminate a subset of projections from the second-order decoder.
The challenge now lies in distributing the projections in a way that maximizes reuse of the second-order decoder.

\subsection{Projection allocation method for RM$(m,3)$ codes}
\begin{figure}[t]
    \centering
        \includegraphics[width=0.49\textwidth]{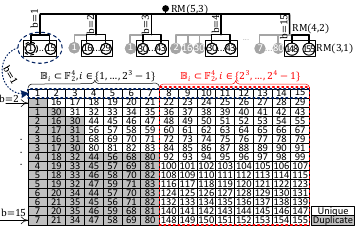}
    
    \caption{Redundancy tree and matrix $\mathbf{R}$ for RM$(5,3)$.}
    \label{fig:UPA:rt}
\end{figure}

Fig.~\ref{fig:UPA:rt} illustrates the unfolded projection levels for an example third-order RM code with $m = 5$ in a tree-like structure, which we call \emph{redundancy tree}.
Since only the first $2^{m-1}-1$ branches (i.e., projections) on the initial level contribute to unique first-order codewords at the innermost level~\cite{Hashemipour2023}, the latter $2^{m-1}$ branches are eliminated from the redundancy tree shown in Fig.~\ref{fig:UPA:rt}. The first-order vectors are represented as leaves on the tree and are numbered from $1$ to $\binom{m}{r-1}_2$. Duplicate vectors share identical numbers.
A matrix $\mathbf{R}$ with size $\left(2^{m-r+2}-1\right) \times \left(2^{m-r+2}-1\right)$ is built from the redundancy tree as shown in Fig.~\ref{fig:UPA:rt}. Each innermost level branch, corresponding to a second-order vector, is illustrated as a row in $\mathbf{R}$. Moreover, the $i$-th column in $\mathbf{R}$ represents the one-dimensional projection of the second-order codewords onto $\mathbb{B}_i=\{0,i\}$. 
The right half of $\mathbf{R}$ always contains unique numbers (white), while the left half contains both unique and duplicate numbers (gray), representing duplicate first-order codewords generated during IPA decoding.

As discussed earlier, the unique projection selection method in IUPA decoding prevents the generation of duplicate projected first-order codewords. 
In a partially-parallel implementation of the IUPA decoder with $G$ second-order decoders, each second-order decoder can perform different sets of second-order projections (i.e., columns in $\mathbf{R}$).
The primary goal is to minimize the hardware required for the IUPA decoder while maintaining a given latency.
To achieve this, we first need to identify all unique numbers (i.e., first-order codewords) generated by the second-order decoders used in the IUPA decoder.
We do this by grouping the rows of $\mathbf{R}$ such that each group contains unique numbers, both within the group and across other groups, while maximizing the overlap in their column indices.
Consequently, rows within a group that share similar columns will utilize the same second-order decoder, reducing hardware requirements.
Moreover, if one PU in each second-order decoder can perform one second-order projection in one clock cycle, we will require $\lceil\frac{\#\text{selected columns}}{\lambda}\rceil$ PUs for each group to finish its assigned second-order projections in $\lambda$ clock cycles. Hence, in a partially-parallel implementation with $G$ second-order decoders, achieving a high degree of column overlap within each group can significantly reduce hardware costs for a given latency. 
The aforementioned problem can be formulated as an integer linear program (ILP) with the objective of minimizing the total number of PUs required for $G$ second-order decoders to achieve a fixed latency $\lambda$.
\subsection{ILP formulation}
\label{sec:ilp}
We use a $(2^{m-1}-1) \times (2^{m-1}-2)$ matrix $\mathbf{D}$ constructed from the left half of $\mathbf{R}$.
The right half of $\mathbf{R}$, which has $2^{m-2}$ columns (i.e., second-order projections), is excluded from $\mathbf{D}$ as it contains only unique first-order vectors. 
These unique first-order vectors require $\lceil\frac{2^{m-2}}{\lambda}\rceil$ dedicated PUs in each second-order decoder to be processed within $\lambda$ cycles, so they do not need to be involved in the ILP formulation.
In $\mathbf{D}$, each number appears three times for RM$(m,3)$ codes.
Hence, for any element ${D}_{jk}$ in $\mathbf{D}$, there exist ${D}_{j'k'}$ and ${D}_{j''k''}$ such that ${D}_{jk} = {D}_{j'k'} = {D}_{j''k''}$, where $(j,k) \ne (j',k') \ne (j'',k'')$.

Along with $\mathbf{D}$, we provide two parameters as inputs to the ILP problem: the target latency $\lambda$ and the number of groups $G$.
Limiting $G$ to a power of two is advantageous because it prevents uneven distribution of rows among the groups, which can complicate the control unit and lead to inefficient resource usage as smaller groups would be idle part of the time.
By making $G$ a power of two, each group contains $\frac{2^{m-1}}{G}$ equally distributed rows, except for one group, which has $\frac{2^{(m-1)}}{G}-1$ rows.
Similarly, limiting the latency of each decoder to $\lambda=2^l, l \in \mathbb{N}$ simplifies the hardware design as  $\frac{2^{m-2}}{\lambda}$ PUs are needed for processing the second-order projections excluded from matrix $\mathbf{D}$ (i.e., right-half of matrix $\mathbf{R}$). 

The objective for the ILP problem is to minimize the total number of PUs required to process the unique numbers in matrix $\mathbf{D}$ among $G$ groups of rows with latency $\lambda$.
To define the ILP constraints, we first introduce the following variables:
\begin{itemize}
\item $x_{ijk}$: A binary variable in $\{0,1\}$ for which $x_{ijk} = 1$ if and only if the $i$-th group processes ${D}_{jk}$.
\item $c_{ik}$: A continuous\footnote{We choose continuous variables because they are computationally cheaper and faster to solve using an ILP solver. All values \(c_{ik} < 1\) and \(r_{ik} < 1\) are rounded down to zero.} variable in $[0, 1]$ for which $c_{ik} = 1$ if the $i$-th group processes the $k$-th column $\mathbf{D}_{\star k}$.
\item $r_{ij}$: A continuous\footnotemark[\value{footnote}] variable in $[0, 1]$ for which $r_{ij} = 1$ if the $i$-th group selects the $j$-th row $\mathbf{D}_{j\star}$ to process.
\item $p_i$: An integer variable in $\mathbb{N}$ indicating the number of PUs allocated for the $i$-th group.
\end{itemize}
In addition, let $\mathcal{G} = \{1, \ldots, G\}$ denote the set of second-order decoders, and $\mathcal{R} = \{1, \ldots, 2^{m-1} - 1\}$ and $\mathcal{C} = \{1, \ldots, 2^{m-2} - 1\}$ represent the sets of rows and columns in the input matrix $\mathbf{D}$, respectively. 
Now, we can formulate the projection allocation problem as follows:
\begingroup
\allowdisplaybreaks
\begin{align}
\text{min}\quad &\label{eq:objFunc} \sum_{i \in \mathcal{G}} p_i\\
\text{s.t.}\quad &\label{eq:phi}\lambda \geq \text{max}_i\left( \frac{\sum_{k \in \mathcal{C}} c_{ik}}{p_i}\right), \;  \forall {i \in \mathcal{G}}\\
&\begin{aligned} \label{eq:uniS}
& \sum_{i \in \mathcal{G}} x_{i jk}+x_{i j'k'}+x_{i j'' k''}=1, \\
& \quad\quad \forall {j, j',j'' \in \mathcal{R}}, \forall {k,k',k'' \in \mathcal{C}}, \\
& \quad\quad D_{j k}=D_{j' k'}=D_{j'' k''}, \\
& \quad\quad \wedge\left(j, k\right) \neq\left(j', k'\right) \neq\left(j'', k''\right),
\end{aligned}\\
& \label{eq:colS}c_{ik} \geq x_{ijk} ,\;  \forall {i \in \mathcal{G}},\forall {j \in \mathcal{R}},\forall {k \in \mathcal{C}}, \\
&\label{eq:rowS} r_{ij} \geq  x_{ijk} ,\;  \forall {i \in \mathcal{G}},\forall {j \in \mathcal{R}},\forall {k \in \mathcal{C}}, \\
&\label{eq:uniRow}\sum_{i \in \mathcal{G}} r_{ij} = 1, \; \forall {j \in \mathcal{R}},\\
&\label{eq:rowUn} \sum_{j \in \mathcal{R}} r_{ij} = \left\lceil \frac{|\mathcal{R}|}{|\mathcal{G}|} \right\rceil ,\; \forall {i \in \mathcal{G}} \wedge i>0,\\
&\label{eq:rowDif} \sum_{j \in \mathcal{R}} r_{ij} = \left\lceil \frac{|\mathcal{R}|}{|\mathcal{G}|} \right\rceil-1 ,\; \forall {i \in \mathcal{G}} \wedge i=0.
\end{align}
\endgroup
Each group $i$ can perform $p_i$ second-order projections (i.e., columns) per clock cycle. 
Constraint \eqref{eq:phi} ensures that every group completes its assigned projections whithin $\lambda$ cycles. 
Constraint \eqref{eq:uniS} ensures that only one out of three copies of each first-order codeword is selected among all groups since  $x_{ijk}$, $x_{ij'k'}$, and $x_{ij''k''}$ are binary variables.
Constraint \eqref{eq:colS} guarantees that if $x_{ijk}=1$, group $i$ must include column $\mathbf{D}_{\star k}$. 
Likewise, if group $i$ selects $D_{jk}$ (i.e., $x_{ijk}=1$), row $\mathbf{D}_{j \star}$ should be assigned to group $i$.
To avoid redundant second-order decoders, constraint \eqref{eq:uniRow} specifies that each row can only belong to one group.
Constraints \eqref{eq:rowUn} and \eqref{eq:rowDif} ensure an even distribution of rows across groups, as previously discussed.

\section{IUPA: Hardware Architecture}
\label{sec:iupa:arch}
\begin{figure}[t]
\centering
\includegraphics[width=0.5\textwidth]{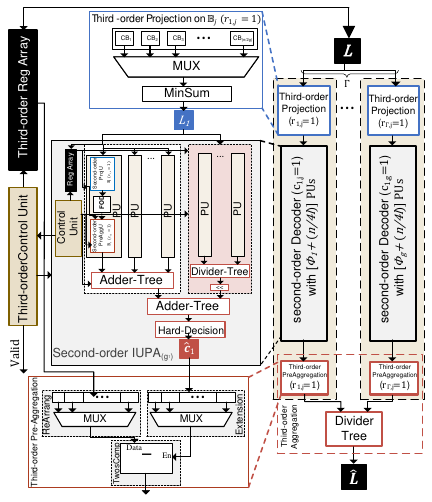}
\caption{Third-order IUPA decoder for one iteration.}
\label{fig:IUPA3rd}
\end{figure} 
In this section, we present a hardware architecture of the soft-input IUPA decoder for third-order RM codes.
Fig.~\ref{fig:IUPA3rd} illustrates the block diagram for the proposed soft-input IUPA decoder, which is designed based on the solution of the ILP in Section~\ref{sec:ilp}.
This decoder consists of $G$ parallel units, each having three primary hardware components: for \textit{third-order projection}, \textit{second-order decoding}, and \textit{third-order pre-aggregation}.
A divider tree is also implemented to determine the final output for the aggregation step. 
Additionally, a register array and a control unit manage the data path and control signals, respectively.
The proposed architecture decodes the input vector $\boldsymbol{L}$ using a partially-parallel approach to adapt to the available resources and desired latency. 

\subsection{Third-order projection}
The third-order projection component for the $i$-th group contains crossbars for $2^{m-1}/G$ different permutations of the input vector $\boldsymbol{L}$ based on the subspaces $\mathbb{B}_j$ selected by the ILP (i.e., $r_{ij}= 1, \forall j \in \mathcal{R}$).
A multiplexer controlled by the control unit selects one permutation at each clock cycle. The output of the multiplexer, which is the permuted input vector of LLRs is directed to the \textit{MinSum} component. The \textit{MinSum} component includes $2^{m-1}$ two-input circuits to perform \eqref{eq:minsum} on every pair of coordinates of the permuted LLR vector. Therefore, at each clock cycle, one projected vector of the received vector $\boldsymbol{L}$ onto the one-dimensional subspace $\mathbb{B}_j$, $r_{ij}=1$ is generated.
   
\subsection{Second-order decoder}
\label{sec:2ndDecImp}
\begin{figure}[t]
    \centering
    \subfloat[]{%
        \label{fig:UPA:ilp2}
        \includegraphics[width=0.24\textwidth]{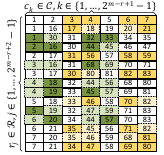}
    }
    \subfloat[]{%
        \label{fig:UPA:ilp4}
        \includegraphics[width=0.24\textwidth]{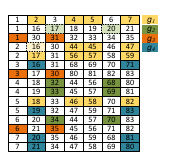}
    }
    
    \caption{Output of ILP optimization for RM$(5,3)$ code with different $G$. (a) $G = 2$ and (b) $G = 4$.}
    \label{fig:combined}
\end{figure}

A partially-parallel second-order decoder is implemented for each group $g_i$ to decode the second-order projected vector  generated by the third-order projection component.
This decoder consists of $p_i$ PUs to handle selected second-order projections for the $i$-th group of rows (i.e., the selected columns $ c_{ik} = 1, \forall k \in \mathcal{C}$).
The ILP output may not always provide a perfect solution. For instance, Fig.~\ref{fig:UPA:ilp2} illustrates an output of the ILP for an RM$(5,3)$ code with $G = 2$ and $ \lambda = 2 $.
In this example, group $g_1$, illustrated in yellow, covers columns $ \{3,4,6,7\} $ and rows $ \{1,2,5,7,10,13,14,15\} $. Although all columns $ \{3,4,6,7\} $ are selected in rows $ \{1,5,15\} $, the other rows assigned to this group do not select all the columns. In such cases, the second-order decoder processes all selected columns for all the second-order vectors dedicated to it.
This may lead to the presence of some duplicate first-order codewords, such as codeword $18$, and the other duplicate codewords highlighted in light colors with dashed lines. 
Nevertheless, this potential overlap simplifies the control unit and is small in practice. 
Notably, reducing the number of rows that are grouped together decreases the number of duplicate codewords. Consequently, a higher number of second-order decoders (i.e., a higher \( G \)) results in fewer duplicate codewords. For instance, in Fig.~\ref{fig:UPA:ilp4}, where $ G = 4 $ and $ \lambda = 2 $, only 3 duplicate second-order codewords are required compared to 25 in the previous case.

The implementation of each PU in the proposed decoder is similar to~\cite{HashemipourTCAS2023}. Within each PU, there exists a second-order projection unit (\textit{ProjU}), one FOD, and a second-order pre-aggregation unit (\textit{PreAggU}).
Based on the ILP output, each group $g_i$ performs second-order projections denoted by $ \mathbb{B}_k $ according to the columns selected (i.e., $c_{ik} = 1 $) for that group. Therefore, each PU dedicated to $g_i$ must accommodate projection circuits for $\lambda$ different second-order projections, such that $\left\lceil \frac{\Sigma_{k\in \mathcal{C}}c_{i,k}}{\lambda}\right\rceil = p_i $.
Consequently, the second-order projection component includes $ \lambda $ different permutations for the projected vector $ \boldsymbol{L}_i $, generating $ \lambda $ first-order codewords in $ \lambda $ clock cycles due to the pipe-lined implementation. Subsequently, the generated first-order codewords are decoded using the FOD, as in \cite{HashemipourTCAS2023}. Following this, the second-order pre-aggregation unit prepares the output of the FOD to compute the average LLR over all second-order vectors processed by group $ g_i $ as defined in \eqref{eq:agg}.
Moreover, in order to achieve the desired latency $\lambda$, each group $g_i$ is also equipped with $ \frac{2^{m-2}}{\lambda}$ extra PUs (highlighted in red in Fig.~\ref{fig:IUPA3rd}) only for the second-order projections corresponding to $ \mathbb{B}_i $ where $ i \in \{2^{m-2}, \dotsc, 2^{m-1}\} $. These projections always result in unique first-order codewords, as illustrated in the right half of the matrix $\mathbf{R}$ in Fig.~\ref{fig:UPA:rt}.

The non-pruned IPA decoder always uses $2^m - 1$ inputs to compute the average, so~\cite{HashemipourTCAS2023} used a divider tree for this purpose. However, when the number of inputs is not a power of two, an alternative method is required for averaging. The objective in the aggregation step \eqref{eq:agg} is to compute the average as the new estimation for the LLR values of the decoded vector for the subsequent iteration.
In the IUPA framework, there are no additional iterations for internal levels. Thus, the goal of the aggregation step for second-order codewords is to determine the sign of the average to finalize decoding at this level and derive the binary decoded vector as the output. Consequently, the average can be replaced by a simple summation since they have the same sign. 
To this end, the output of the PUs for the initial segment of group $g_i$, which processes the projections selected from the ILP output, are inserted into a full-precision adder tree as depicted in Fig.~\ref{fig:IUPA3rd}.

However, the output of the PUs for the second segment of group $g_i$ (highighted in red) is directed to a divider tree initially to compute the average of $ 2^{m-2-l}$ decoded codewords. The divider tree is selected due to its low hardware requirement compared to a full-precision adder tree.
To ensure that the output of the divider tree has the same numerical impact on the final average of this group as the output of the adder tree, we need to multiply it by $2^{m-2-l}$. This is achieved by appending $m-2-l$ zero bits to the right of the  output of divider tree.
To derive the final output for each second-order decoder, we must accumulate the results from the first and second segments using the adder tree depicted in Fig.~\ref{fig:IUPA3rd}. Finally, a hard-decision is made based on the output of the adder for the $i$-th group to estimate the binary codeword $ \hat{\boldsymbol{c}}_i $ for the next level.
\subsection{Third-order pre-aggregation}
The third-order pre-aggregation unit follows the design of the second-order pre-aggregation unit~\cite{HashemipourTCAS2023}. Similar to the projection unit, the crossbars described in Section \ref{sec:ipa} are designed to perform appropriate permutations based on the subspaces $ \mathbb{B}_j $ chosen for group $ g_i $ from the ILP result (i.e., $ r_{ij} = 1 $).

\subsection{Divider tree}
In IUPA decoding for third-order RM codes, there are always $2^{m-2}-1$ third-order pre-aggregated vectors. In our proposed IUPA decoder, we introduce an extra dummy all-zero vector generated from running the second-order decoder for the additional row added to the output of the ILP for the group with $2^{m-1-\log_2(G)}-1$ rows.
Consequently, there are $2^{m-1}$ pre-aggregated vectors to average, which we can average with the divider tree used in the IPA decoder, shown in Fig.~\ref{fig:SIPABD}.
Considering that we require at least two groups for the third-order IUPA decoder, more than one input is available at every clock cycle for the divider. Additionally, based on the constraints defined in Section~\ref{sec:ilp} for the number of groups, the divider tree  receives a power-of-two number of inputs at each clock cycle. Therefore, we can replace the shift registers depicted in Fig.~\ref{fig:SIPABD} with standard registers for the first $\log_2(G)$ levels of the divider tree. Thus, we have $2^{\log_2(G)-l}$ two-input standard registers at first $\log_2(G)$ levels. These registers are activated with the valid output of the \textit{pre-aggregation} unit.
We maintain the sequential part of the divider tree operating with shift registers for the last $m-\log_2(G)$ levels of the dividing. The shift register placed in the first level of the sequential part, which is the ($m-\log_2(G)$)-th level, is activated with the valid signal of the \textit{pre-aggregation} unit delayed by $\log_2(G)$ clock cycles. The activation signals for the remaining shift registers are generated by the control unit.
\subsection{Register array}
In Fig.~\ref{fig:IUPA3rd}, each second-order decoder has a register array responsible for storing the LLR values of the second-order codewords entering that decoder. This array is crucial for the pre-aggregation step which requires the LLRs of second-order codewords as shown in \eqref{eq:agg}. In a pipelined structure, registers are necessary to preserve this vector across stages for each PU. To minimize area, all PUs within a group share a register array with read and write operations managed by a central control unit. As a result, the requirement for transporting this vector across pipeline stages is eliminated, and this saves multiple registers. The depth of this array is:
\begin{equation} \label{eq:regArr}
D_{\text{RA2nd}} = \bigg\lceil{\frac{l_{\text{agg2nd}}}{\lambda}}\bigg\rceil+1,
\end{equation}
where $l_{\text{agg2nd}}$ is the number of pipeline stages from generating a new second-order vector until the pre-aggregation unit in the PU. Similarly, a register array is essential for the third-order level due to the same requirement. The depth of this array is:
\begin{equation} \label{eq:regArr3rs}
D_{\text{RA3rd}} = \bigg\lceil{\frac{l_{\text{agg3rd}}}{\lambda \times n/2 \times G}}\bigg\rceil+1,
\end{equation}
where $l_{\text{agg3rd}}$ is the length of pipeline stages from the insertion of a new vector until it reaches the third-order pre-aggregation unit.
For most configurations, $D_{\text{RA3rd}} = 2$, as the denominator of the fraction in \eqref{eq:regArr3rs} exceeds its nominator.

\subsection{Throughput and latency}
In the proposed third-order IUPA decoder configured with the output of the ILP model with input parameters $G$ and $\lambda$, a new codeword with a block length of $n$ can be inserted every $\frac{n \times \lambda}{2G}$ clock cycles. Therefore, the throughput is:
\begin{equation} \label{eq:thr}
{\text{Throughput}_{\text{Mbps}}} = \frac{2G \times \text{Frequency}}{\lambda},
\end{equation}
where the frequency is given in MHz.
The latency of our decoder in clock cycles is given by:
\begin{align}
\label{eq:latencyr}
\scalemath{0.8}{
{t}{(m,3)} = t_{\text{proj}} + t_{\text{group}} + t_{\text{preAgg}} + \left\lceil\frac{\left(2^{m-1}-1\right)\lambda}{G}\right\rceil + m,}
\end{align} 
where $t_{\text{proj}}=1$ and $t_{\text{preAgg}}=1$ are the latencies of the projection and pre-aggregation unit, respectively. ${{t}_{\text{group}}}$ is the latency of the group or the second-order decoder:
\begin{align}
\label{eq:grlatency}
{t}_{\text{group}} = \left(t_{\text{proj}}+t_{\text{FOD}}+t_{\text{preAgg}}\right)+\lambda+1,
\end{align}
where $t_{\text{FOD}}$ is the latency of first-order decoder and equals 3 or 4, depending on the block-length of the target RM code~\cite{HashemipourTCAS2023}.

\section{CPA: Hardware Architecture}
\label{sec:CPA_imp}
\begin{figure}[t]
\centering
\includegraphics[width=0.5\textwidth]{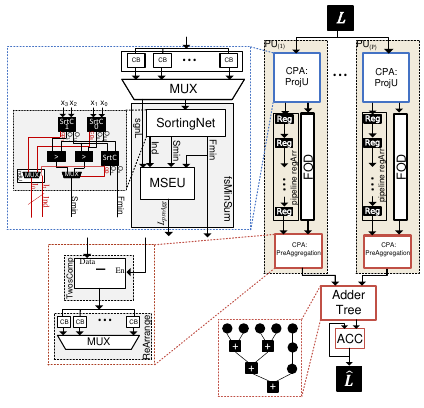}
\caption{Block diagram of CPA decoder for one iteration.}
\label{fig:CPA}
\end{figure} 

If we consider the number of first-order decodings as a proxy for the computational complexity of IUPA decoding, the complexity of CPA decoding is identical to IUPA decoding. However, the question about which decoder is more efficient in terms of hardware implementation was left open by \cite{Hashemipour2023}. 
To this end, in this section, we propose an efficient CPA decoding hardware architecture in order to compare CPA and IUPA.
\subsection{Architecture}
As discussed in Section~\ref{Sec:CPA}, the CPA decoder projects the received LLR vector $\boldsymbol{L}$ onto $(r-1)$-dimensional subspaces to construct first-order codewords. Therefore, it involves a single projection level, a first-order decoding step, and subsequent aggregation. Fig.~\ref{fig:CPA} illustrates our proposed CPA decoder. The architecture follows the design of IUPA and IPA decoders, including $p$ PUs, each with a projection unit, FOD, and a pre-aggregation unit. Similar to the IPA and IUPA decoders, we impose a constraint on the number of PUs, limiting $p$ to be a divisor of ${{\binom{m}{r-1}}_2}$ to simplify the control unit.

\subsubsection{PU}
\label{Sec:CPA:PU}
Similar to the IUPA decoder, the projection unit in each PU of the CPA decoder consists of modules for reordering the coordinates of the input LLR vector and calculating the MS operation. The reordering module uses $\tfrac{{\binom{m}{r-1}}_2}{p}$ crossbars to rearrange the coordinates of the input vector $\boldsymbol{L}$. This step groups the coordinates according to their corresponding $(r-1)$-dimensional subspaces, with each group containing $2^{r-1}$ coordinates.
As shown in Fig.~\ref{fig:CPA}, the MS operation is implemented in the \textit{fsMinSum} component. 
This module carries out the MS calculation described in~\eqref{eq:minsum}, determining the minimum absolute value along with its index, the sign, as well as the second minimum absolute value.
To perform the sorting process, we use the \textit{SortingNet} sub-component, which is based on a pruned sorting network called tree structure (TS) proposed in \cite{FShwMIN}.
This component finds the first and second minimum, denoted by $f_{\min}$ and $s_{\min}$, respectively, among the absolute value of the inputs of the \textit{fsMinSum} component as well as the index of the first absolute minimum, denoted by $I$.

The first minimum found by the \textit{SortingNet} component is sent to the FOD component, while the second minimum goes to the MS extension unit \textit{MSEU} to prepare the input for the pre-aggregation step.
To estimate the coordinate $\hat{L}(z)$ during the aggregation step as described in \eqref{eq:agg:cpa}, we perform an MS operation on the vector $\boldsymbol{L}$, considering the coordinates indexed by the elements of the coset $T$, except for the one indexed by $z$.
This is similar to the projection step in \eqref{eq:proj}, but excluding the coordinate indexed by $z$.
The MS operation in \eqref{eq:agg:cpa} can be broken into two parts:
\begin{equation}
\label{eq:LpreAgg}
\begin{aligned}
\min_{z_{j}\in T-z}\lbrace|L(z_j)|\rbrace = \begin{cases} |f_{\min}|, & \text { if } I \neq z,  \\
|s_{\min}|, & \text { if } I = z\end{cases}. \\
\end{aligned}
\end{equation}

\begin{equation}
\label{eq:LpreaggSign}
\scalemath{0.85}{
\prod_{z_{j}\in T-z} \text{sign}\left(L\left(z_{j}\right)\right) =  \left(\prod_{z_{j}\in T} \text{sign}\left(L\left(z_{j}\right)\right)\right) \oplus \text{sign}\left(L\left(z\right)\right)
}
\end{equation}
A multiplexer selects between $f_{\min}$ and $s_{\min}$, using $I$ as the selector to implement \eqref{eq:LpreAgg}. Additionally, equation \eqref{eq:LpreaggSign} can be implemented with an XOR gate with two inputs, the sign output from \textit{fsMinSum} component and the binary vector $L_{\text{sgn}}$, containing the sign bits from the coordinates of the input LLR vector $\boldsymbol{L}$.
Considering the above, we can rewrite \eqref{eq:agg:cpa} as:
\begin{equation}\label{eq:agg:cpa:simp}
\scalemath{0.85}{
{\hat{L}}(z) = \frac{1}{n_p} \sum_{i=1}^{n_P} {-1^{\hat{c}_{/ \mathbb{B}_i}([T])} 
\boldsymbol{L}_{\text{preAgg}}\left(z\right)}
},
\end{equation}
where $\boldsymbol{L}_{\text{preAgg}}$ is a vector of LLRs, for which the absolute values are calculated based on \eqref{eq:LpreAgg}, and the signs are determined based on \eqref{eq:LpreaggSign}.
We need to keep $\boldsymbol{L}_{\text{preAgg}}$ for the pre-aggregation unit during the pipeline stages of FOD, which typically requires $3$ to $4$ clock cycles depending on the code blocklength. 
As a result, the pipeline register array consisting of three or four registers, where the output of each register serves as the input for the subsequent register as depicted in Fig.~\ref{fig:CPA}.

To implement the aggregation step, we follow the design principles of the IPA and IUPA decoders, which break this step into two sub-steps. In the first sub-step, the value inside the summation in \eqref{eq:agg:cpa:simp} is calculated. This is done by using the decoded binary vector ${\hat{c}_{/ \mathbb{B}_i}}$, which is the output of the FOD, along with the $\boldsymbol{L}_{\text{preAgg}}$ vector in the pre-aggregation unit.
As explained in Section~\ref{sec:ipa}, the term ${-1^{\hat{c}_{/ \mathbb{B}_i}([T])} 
\boldsymbol{L}_{\text{preAgg}}\left(z\right)}$ can be implemented with a component calculating the two's complement of its input. The input to this unit is the $\boldsymbol{L}_{\text{preAgg}}$ vector, with the enable port controlled by ${\hat{c}_{/ \mathbb{B}_i}}$. This functionality is managed by the \textit{TwosCom} subunit of the pre-aggregation unit. Additionally, because the coordinates in $\boldsymbol{L}_{\text{preAgg}}$ are reordered in the projection unit, we need to restore them to their original order using the \textit{ReArrange} unit.

\subsubsection{Average}
\label{sec:cpa:average}
To finalize the aggregation step, we take the average of $\binom{m}{r-1}_2$ estimations, which requires division.
However, since the final decoded output is determined by the sign of the average, we can simply add up all estimated vectors instead. 
Since there are $p$ PUs, we can insert $p$ inputs into the adder at each clock cycle. This structure allows us to perform the additions using an adder tree, as depicted in Fig.~\ref{fig:CPA}. The output of the adder tree feeds into an accumulator, calculating the final summation every ${\binom{m}{r-1}_2}/{p}$ clock cycles.
Both the adder tree and the accumulator can be implemented with either full-precision or saturated adders with configurable precision.\footnote{This choice is discussed further in Section~\ref{Sec:res:fp}.}

Fig.~\ref{fig:CPA} illustrates the hardware for one iteration of CPA decoding. To implement additional iterations, this architecture can be replicated as needed, with the output of each iteration being the input to the next iteration. It is essential to maintain a limited bit-width for the input of each subsequent iteration. If each LLR is represented by $Q$ bits, then the output from the accumulator at the end of each iteration is saturated to fit within these $Q$ bits for the next iteration.
\subsection{Throughput and latency}
In our CPA decoder with $p$ PUs, a new codeword of length $n$ can be inserted into the pipeline every $\frac{{\binom{m}{r-1}}_2}{p}$ clock cycles.
Hence, the throughput of the CPA decoder is:
\begin{equation} \label{eq:CPAthr}
{\text{Throughput}_{\text{Mbps}}} = \frac{p \times \text{Frequency}}{\binom{m}{r-1}_2}\times n,
\end{equation}
where the frequency is given in MHz. Furthermore, the latency of one iteration of the CPA decoder can be expressed as:
\begin{align}
\label{eq:CPAlatency}
\scalemath{0.8}{
{t_{\text{CPA}}} = 1+ t_{\text{proj}} + t_{\text{FOD}} + t_{\text{preAgg}} + t_{\text{Add}} + \frac{\binom{m}{r-1}_2}{p} }.
\end{align}
Similar to IUPA, $t_{\text{proj}} = t_{\text{preAgg}} = 1$, and $t_{\text{FOD}}$ is 3 or 4 clock cycles, depending on the code block length. $t_{\text{Add}}$ represents the latency of the parallel adder tree for $p$ inputs and $\text{log}_2p$ stages, and its value is 1 or 2, depending on $p$.

\section{Simulation and Implementation Results}
\label{sec:result}
In this section, we present the simulation results for the frame error rate (FER) of the proposed IUPA and CPA decoders, followed by the synthesis results.
\subsection{Simulation results}
\label{sec:sim_res}

In this section, we present the results of the ILP described in Section~\ref{sec:ilp} for various parameters. We analyze different number of groups and delays to provide a detailed trade-off between latency and resource usage.
Additionally, we compare the simulation results of the ILP-aided IUPA decoder with the IPA and CPA decoding methods. We also explore different numbers of iterations and quantization bit-widths to find an efficient configuration for the hardware implementation.

\subsubsection{Latency-resource usage trade-offs}
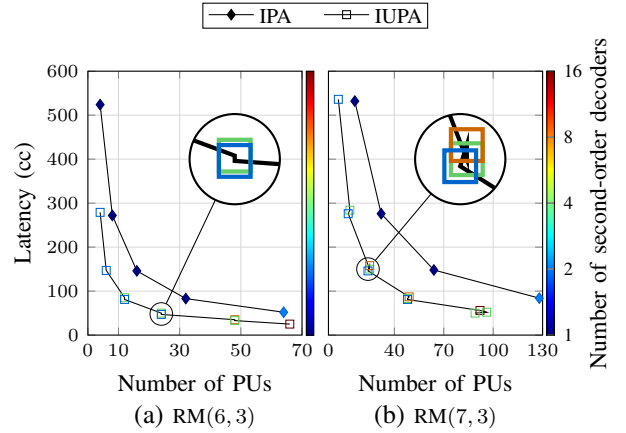
\begin{figure}[t]
    \centering
    \begin{tikzpicture}[spy using outlines={magnification=4,circle,size=1.2cm, black,connect spies}]
		\begin{groupplot}[group style={group name=ILPpareto, group size= 2 by 1, horizontal sep=10pt, vertical sep=5pt},
			footnotesize,
			height=.575\columnwidth,  width=.50\columnwidth,
			xlabel= {Number of PUs},
			grid=both, grid style={gray!30},
			/pgfplots/table/ignore chars={|},
			]

			\nextgroupplot[ylabel= Latency (cc), ytick pos=left, y label style={at={(axis description cs:-0.20,.5)},anchor=south},
			colormap/jet,
    		point meta min=0,
    		point meta max=4,
    		colorbar,
    		colorbar style={
	        width=0.1cm,
	        yticklabels=\empty,
	        ytick = \empty,
        	at={(1.02,0.5)}, anchor=west 
    		},
    		xmin=0, xmax=70, xtick={0,10,30,50,70},
    		ymin=0, ymax=600,
    		]
    		
\coordinate (pl1) at (axis cs: 24,48);
\coordinate (pl2) at (axis cs: 48,400);
\spy on (pl1) in node[fill=white] at (pl2);

\addplot[scatter, scatter src=explicit, mark=diamond*, mark size=2,colorbar=false] table[x=numProcessors, y=latency, meta=log2numGroups,colorbar= false] {
numProcessors  latency numGroups  log2numGroups
4				524		1			0
8				272     1			0
16				146     1			0	
32				83      1			0
64    			52	    2			1
};\label{pareto:IPA}

\addplot[scatter,  scatter src=explicit, mark= square, mark size=1.5] table[x=numProcessors, y=latency, meta=log2numGroups] {
numProcessors  latency numGroups   log2numGroups
4				279		2			1
6      			147 	2			1    
12				85		4			2
12      		81		2			1
24				50		4			2
24				47		2			1
48				35		8			3
48				33		4			2
66		    	25		16			4
};
         
\coordinate (top) at (rel axis cs:0,1);

\nextgroupplot[yticklabels=\empty,
			   colormap/jet,
    		point meta min=0,
    		point meta max=4,
    		colorbar,
    		colorbar style={
    		ylabel=Number of second-order decoders,
    		ylabel style={
        	yshift=4pt, 
    		},
	        width=0.1cm,
	        ytick = {0,1,2,3,4},
	        yticklabels = {1,2,4,8,16},
        	at={(1.02,0.5)}, anchor=west 
    		},
    		xmin=0, xmax=130, xtick={0,40,70,100,130},
    		ymin=0, ymax=600,
    		]

\coordinate (p21) at (axis cs: 24,150);
\coordinate (p22) at (axis cs: 80,400);
\spy on (p21) in node[fill=white] at (p22);

\coordinate (bot) at (rel axis cs:1,0);

\addplot[scatter,  scatter src=explicit, mark=diamond*, mark size=2,colorbar=false] table[x=numProcessors, y=latency, meta=log2numGroups,colorbar= false] {
numProcessors  latency numGroups	log2numGroups
16				532      1				0
32				276      1				0
64    			148		 1				0
128   			84		 2				1
};

\addplot[scatter,  scatter src=explicit, mark= square, mark size=1.5] table[x=numProcessors, y=latency, meta=log2numGroups] {
numProcessors 	latency	 numGroups		log2numGroups
     6          536			2				1	
    13   		284			4				2
    12   		276			2				1
    25   		150			4    		    2
    25   		158			8				3     
    24   		146			2				1
    49    		87			8				3
    48    		83			4				2    
    48    		81			2				1
    92    		56			16				4
    96    		52			4    		    2
    89    		50			4				2
};\label{pareto:IUPA}

		\end{groupplot}
		\node[below = 0.8cm of ILPpareto c1r1.south] {(a) \footnotesize RM$(6,3)$ };
		\node[below = 0.8cm of ILPpareto c2r1.south] {(b) \footnotesize RM$(7,3)$};
		\path (top|-current bounding box.north) -- coordinate(legendpos) (bot|-current bounding box.north);
		\matrix[
		matrix of nodes,
		anchor=south,
		draw,
		inner sep=0.1em,
		draw,
		column 1/.style={anchor=base west},
    	column 2/.style={anchor=base west},
    	column 3/.style={anchor=base west},
    	column 4/.style={anchor=base west},
		]at(legendpos)
		{
			\ref{pareto:IPA}& \footnotesize IPA &[10pt]
			\ref{pareto:IUPA}& \footnotesize IUPA  \\
			};
	\end{tikzpicture}
    \caption{Latency-resource usage trade-offs for IPA and ILP-aided IUPA decoding for (a) RM$(6,3)$ and (b) RM$(7,3)$}
    \label{fig:paretoRM73}
\end{figure}
Fig.~\ref{fig:paretoRM73} illustrates the relationship between the per-iteration latency and the resource usage (measured using the number of PUs) for the IPA and ILP-aided IUPA decoder applied to RM$(6,3)$ and RM$(7,3)$ codes. For IUPA, each color indicates the number of second-order decoders (i.e., group $G$). 
To solve the ILP for the IUPA decoder, we use a cluster with 64 CPU cores at a clock speed of 2.6 GHz and a total memory of 512 GB, using Gurobipy (v10.0.2) as the ILP solver. 
The execution of the solver was set to stop if the solver did not find a new feasible solution with improved number of PUs within six hours.

The results show that the number of PUs required for the IUPA decoder for third-order RM codes is nearly one-third of that needed for the IPA decoder at similar latencies, which aligns with our initial expectations discussed in Section~\ref{sec:iupa}.
Based on~\eqref{NFOD}, one-third of first-order codewords are generated in IUPA decoding for RM$(m,3)$ codes compared to IPA. As shown in Fig.~\ref{fig:paretoRM73}, we observe a similar gain with our proposed hardware-friendly approach for even distribution of the unique projections.
Some points on the Pareto front for the IUPA decoder overlap, indicating identical latencies and number of PUs, but a different number of groups. It is essential to consider both the number of PUs and the number of second-order decoders in hardware implementations since both impact resource usage. As discussed further in Section~\ref{sec:res:synth}, configurations with fewer groups are smaller than those with more groups, even when the number of PUs is the same. 
This is because each group requires a separate second-order decoder, along with its control unit and register arrays. 
Moreover, achieving extremely low latency requires high parallelism, leading to a higher number of groups.
Therefore, simply decreasing the input parameter $\lambda$ in the ILP formulation is insufficient for achieving very low latency, since the total latency measured in number of clock cycles is also influenced by the number of groups.

\subsubsection{IUPA vs. CPA vs. IPA}
\begin{figure}[t]
	\centering
	\begin{tikzpicture}[spy using outlines={magnification=3,circle,size=1.25cm, black,connect spies}]
		\begin{groupplot}[group style={group name=fer_queries, group size= 2 by 1, horizontal sep=10pt, vertical sep=5pt},
			footnotesize,
			height=.65\columnwidth,  width=.575\columnwidth,
			xlabel=Eb\slash No dB,
			ymode=log,
			tick align=inside,
			grid=both, grid style={gray!30},
			/pgfplots/table/ignore chars={|},
			]

			\nextgroupplot[ylabel= FER, ytick pos=left, y label style={at={(axis description cs:-0.20,.5)},anchor=south},ymin=1e-5, ymax = 1e-1,xmin=3.25, xmax=5.25, xtick={3,3.5,4,4.5,5}]

\coordinate (pl1) at (axis cs: 4.25,1e-3);
\coordinate (pl2) at (axis cs: 3.75,1e-4);
\spy on (pl1) in node[fill=white] at (pl2);

\addplot[ color=blue ,mark=square* ] coordinates {
( 3.25, 0.01872063)
( 3.50, 0.00981402)
( 3.75, 0.00572118)
( 4.00, 0.00269307)
( 4.25, 0.00122413)
( 4.50, 0.00059900)
( 4.75, 0.00024400)
( 5.00, 0.00010700)
( 5.25, 0.00002400)
( 5.50, 0.00000700)
};\label{gp:IUPA}

\addplot[ color=black ,mark=diamond* ] coordinates {
( 3.25, 0.01849968)
( 3.50, 0.00973672)
( 3.75, 0.00571350)
( 4.00, 0.00264934)
( 4.25, 0.00120013)
( 4.50, 0.00058000)
( 4.75, 0.00025600)
( 5.00, 0.00010400)
( 5.25, 0.00002600)
( 5.50, 0.00000600)
};\label{gp:IPA}

\addplot[ color=red ,mark=triangle* ] coordinates {
( 3.25, 0.01876736)
( 3.50, 0.00983275)
( 3.75, 0.00585329)
( 4.00, 0.00273787)
( 4.25, 0.00123858)
( 4.50, 0.00061300)
( 4.75, 0.00025600)
( 5.00, 0.00011700)
( 5.25, 0.00002900)
( 5.50, 0.00000800)
};\label{gp:CPA}

\addplot[ color=Paired-3 ,mark=square ] coordinates {
( 3.25, 0.01847300)
( 3.50, 0.01022000)
( 3.75, 0.00543100)
( 4.00, 0.00273700)
( 4.25, 0.00120300)
( 4.50, 0.00059500)
( 4.75, 0.00024700)
( 5.00, 0.00010900)
( 5.25, 0.00002800)
( 5.50, 0.00000700)
};\label{gp:IUPAG2L4}

\addplot[ color= Paired-7,mark=square ] coordinates {
( 3.25, 0.01854100)
( 3.50, 0.01020700)
( 3.75, 0.00543100)
( 4.00, 0.00274900)
( 4.25, 0.00120200)
( 4.50, 0.00059000)
( 4.75, 0.00025700)
( 5.00, 0.00010300)
( 5.25, 0.00002800)
( 5.50, 0.00000700)
};;\label{gp:IUPAG4L4}

			\coordinate (top) at (rel axis cs:0,1);

\nextgroupplot[yticklabels=\empty,ymin=1e-5, ymax = 1e-1,xmin=2, xmax=4, xtick={2,2.5,3,3.5,4}]

\coordinate (bot) at (rel axis cs:1,0);

\coordinate (p211) at (axis cs: 3.25,1e-3);
\coordinate (p212) at (axis cs: 2.6,1e-4);
\spy on (p211) in node[fill=white] at (p212);

\addplot[ color=blue ,mark=square* ] coordinates {
( 2.00, 0.05828865)
( 2.25, 0.03017502)
( 2.50, 0.01629036)
( 2.75, 0.00736000)
( 3.00, 0.00295000)
( 3.25, 0.00116000)
( 3.50, 0.00046000)
( 3.75, 0.00015900)
( 4.00, 0.00005100)
};

\addplot[ color=black ,mark=diamond* ] coordinates {
( 2.00, 0.05568000)
( 2.25, 0.03020000)
( 2.50, 0.01570000)
( 2.75, 0.00724000)
( 3.00, 0.00287000)
( 3.25, 0.00113000)
( 3.50, 0.00045000)
( 3.75, 0.00015900)
( 4.00, 0.00005200)
};

\addplot[ color=red ,mark=triangle* ] coordinates {
( 2.00, 0.05415358)
( 2.25, 0.02888838)
( 2.50, 0.01527370)
( 2.75, 0.00694000)
( 3.00, 0.00286000)
( 3.25, 0.00111000)
( 3.50, 0.00045000)
( 3.75, 0.000158)
( 4.00, 0.0000500)
};

\addplot[ color=Paired-3 ,mark=square ] coordinates {
( 2.00, 0.05670000)
( 2.25, 0.03069000)
( 2.50, 0.01589000)
( 2.75, 0.00733000)
( 3.00, 0.00306000)
( 3.25, 0.00118000)
( 3.50, 0.00046000)
( 3.75, 0.00017000)
( 4.00, 0.00006000)
};

\addplot[ color=Paired-9 ,mark=square ] coordinates {
( 2.00, 0.05657000)
( 2.25, 0.03021000)
( 2.50, 0.01575000)
( 2.75, 0.00722000)
( 3.00, 0.00299000)
( 3.25, 0.00116000)
( 3.50, 0.00046000)
( 3.75, 0.00017000)
( 4.00, 0.00006000)
};\label{gp:IUPAG4L2}

		\end{groupplot}
		\node[below = 0.8cm of fer_queries c1r1.south] { (a) \footnotesize RM$(6,3)$ };
		\node[below = 0.8cm of fer_queries c2r1.south] {(b) \footnotesize RM$(7,3)$};
		\path (top|-current bounding box.north) -- coordinate(legendpos) (bot|-current bounding box.north);
		\matrix[
		matrix of nodes,
		anchor=south,
		draw,
		inner sep=0.1em,
		draw,
		column 1/.style={anchor=base west},
    	column 2/.style={anchor=base west},
    	column 3/.style={anchor=base west},
    	column 4/.style={anchor=base west},
		]at(legendpos)
		{
			\ref{gp:IPA}& \scriptsize IPA &[3pt]
			\ref{gp:CPA}& \scriptsize CPA  &[3pt]
			\ref{gp:IUPA}& \scriptsize IUPA \\
			\ref{gp:IUPAG2L4}& \scriptsize IUPA-ILP$(2,4)$&[3pt]
			\ref{gp:IUPAG4L4}& \scriptsize IUPA-ILP$(4,4)$ &[3pt]
			\ref{gp:IUPAG4L2}& \scriptsize IUPA-ILP$(4,2)$\\
			};
	\end{tikzpicture}
	\caption{Comparison of different flavours of IPA, IUPA, and CPA for Reed-Muller codes over AWGN channel.}
	\label{fig:iupa_cpa}
\end{figure}
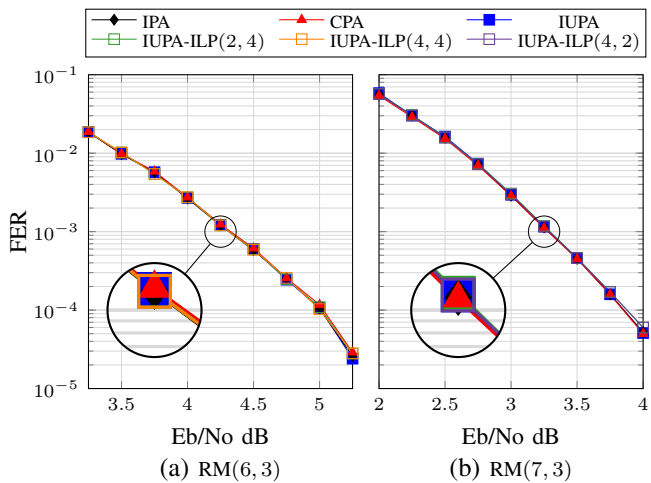
Fig.~\ref{fig:iupa_cpa} shows the FER for the floating-point IPA, CPA, and IUPA decoders for RM$(6,3)$ and RM$(7,3)$ codes. For all decoders, the maximum number of iterations $N_{\max}$ is set to $\lceil{\frac{m}{2}}\rceil$~\cite{Ye2020}. As shown for both codes, CPA and IUPA decoding exhibit the same error-correcting performance, which is comparable to the IPA decoding without pruning. Additionally, the ILP-aided IUPA with different values for $\left(\lambda,G\right)$, which result in varying numbers of duplicate first-order codewords as explained in Section~\ref{sec:2ndDecImp}, has the same FER as the standard IUPA with only unique first-order codewords.
This result is expected because the duplicate first-order codewords do not add new information and thus neither improve nor degrade the performance.

\subsubsection{Number of iterations}
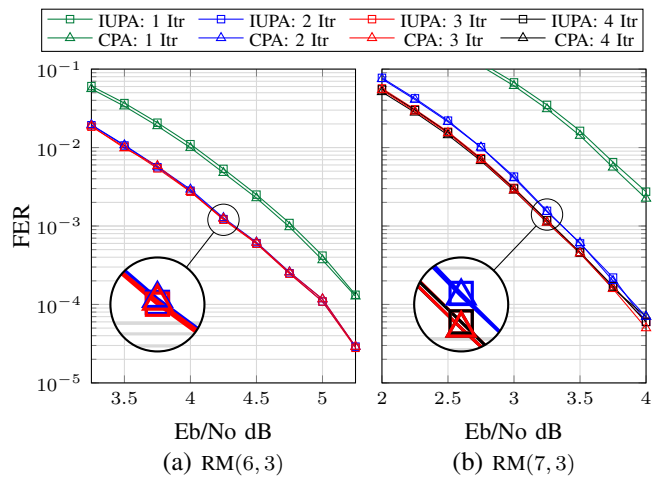
\begin{figure}[t]
	\centering
	\begin{tikzpicture}[spy using outlines={magnification=3,circle,size=1.25cm, black,connect spies}]
		\begin{groupplot}[group style={group name=fer_queries, group size= 2 by 1, horizontal sep=10pt, vertical sep=5pt},
			footnotesize,
			height=.65\columnwidth,  width=.575\columnwidth,
			xlabel=Eb\slash No dB,
			ymode=log,
			tick align=inside,
			grid=both, grid style={gray!30},
			/pgfplots/table/ignore chars={|},
			]

\nextgroupplot[ylabel= FER, ytick pos=left, y label style={at={(axis description cs:-0.20,.5)},anchor=south},ymin=1e-5, ymax = 1e-1,xmin=3.25, xmax=5.25, xtick={3,3.5,4,4.5,5,5.5}]
			
\coordinate (pl1) at (axis cs: 4.25,1.2e-3);
\coordinate (pl2) at (axis cs: 3.75,1e-4);
\spy on (pl1) in node[fill=white] at (pl2);

\addplot[ color=dgreen ,mark=square,mark options={scale=0.7} ] coordinates {
( 3.25, 0.06069700)
( 3.50, 0.03688200)
( 3.75, 0.02073300)
( 4.00, 0.01103700)
( 4.25, 0.00534900)
( 4.50, 0.00251100)
( 4.75, 0.00109200)
( 5.00, 0.00041900)
( 5.25, 0.00013300)
( 5.50, 0.00004000)
};
%
\addplot[ color=blue ,mark=square,mark options={scale=0.7} ] coordinates {
( 3.25, 0.01923400)
( 3.50, 0.01062000)
( 3.75, 0.00560700)
( 4.00, 0.00281800)
( 4.25, 0.00122500)
( 4.50, 0.00060400)
( 4.75, 0.00025000)
( 5.00, 0.00010900)
( 5.25, 0.00002900)
( 5.50, 0.00000700)
};
%
\addplot[ color=red ,mark=square,mark options={scale=0.7} ] coordinates {
( 3.25, 0.01847300)
( 3.50, 0.01022000)
( 3.75, 0.00543100)
( 4.00, 0.00273700)
( 4.25, 0.00120300)
( 4.50, 0.00059500)
( 4.75, 0.00024700)
( 5.00, 0.00010900)
( 5.25, 0.00002800)
( 5.50, 0.00000700)
};
%
\addplot[ color=dgreen ,mark=triangle ] coordinates {
( 3.25, 0.05592900)
( 3.50, 0.03362900)
( 3.75, 0.01882800)
( 4.00, 0.00994700)
( 4.25, 0.00477000)
( 4.50, 0.00227300)
( 4.75, 0.00097100)
( 5.00, 0.00036800)
( 5.25, 0.00012700)
( 5.50, 0.00002900)
};
%
\addplot[ color=blue ,mark=triangle ] coordinates {
( 3.25, 0.01952400)
( 3.50, 0.01075800)
( 3.75, 0.00572700)
( 4.00, 0.00293000)
( 4.25, 0.00127200)
( 4.50, 0.00062100)
( 4.75, 0.00025900)
( 5.00, 0.00011700)
( 5.25, 0.00002900)
( 5.50, 0.00000800)
};
%
\addplot[ color=red ,mark=triangle ] coordinates {
( 3.25, 0.01876736)
( 3.50, 0.00983275)
( 3.75, 0.00585329)
( 4.00, 0.00273787)
( 4.25, 0.00123858)
( 4.50, 0.00061300)
( 4.75, 0.00025600)
( 5.00, 0.00011700)
( 5.25, 0.00002900)
( 5.50, 0.00000800)
};

			\coordinate (top) at (rel axis cs:-0.2,1);
\nextgroupplot[yticklabels=\empty,ymin=1e-5, ymax = 1e-1,xmin=2, xmax=4, xtick={2,2.5,3,3.5,4}]

\coordinate (bot) at (rel axis cs:1,0);

\coordinate (p211) at (axis cs: 3.25,1.4e-3);
\coordinate (p212) at (axis cs: 2.6,1e-4);
\spy on (p211) in node[fill=white] at (p212);

\addplot[ color=dgreen ,mark=square,mark options={scale=0.7} ] coordinates {
( 2.00, 0.38640000)
( 2.25, 0.28198000)
( 2.50, 0.19013000)
( 2.75, 0.12054000)
( 3.00, 0.06769000)
( 3.25, 0.03497000)
( 3.50, 0.01631000)
( 3.75, 0.00652000)
( 4.00, 0.00274000)
};\label{gp:IUPAitr1}

\addplot[ color=blue ,mark=square,mark options={scale=0.7} ] coordinates {
( 2.00, 0.07732000)
( 2.25, 0.04237000)
( 2.50, 0.02201000)
( 2.75, 0.01012000)
( 3.00, 0.00426000)
( 3.25, 0.00156000)
( 3.50, 0.00060000)
( 3.75, 0.00022000)
( 4.00, 0.00006000)
};\label{gp:IUPAitr2}

\addplot[ color=red ,mark=square,mark options={scale=0.7} ] coordinates {
( 2.00, 0.05670000)
( 2.25, 0.03069000)
( 2.50, 0.01589000)
( 2.75, 0.00733000)
( 3.00, 0.00306000)
( 3.25, 0.00118000)
( 3.50, 0.00046000)
( 3.75, 0.00017000)
( 4.00, 0.00006000)
};\label{gp:IUPAitr3}

\addplot[ color=black ,mark=square,mark options={scale=0.7} ] coordinates {
( 2.00, 0.05505000)
( 2.25, 0.03003000)
( 2.50, 0.01553000)
( 2.75, 0.00709000)
( 3.00, 0.00299000)
( 3.25, 0.00118000)
( 3.50, 0.00046000)
( 3.75, 0.00017000)
( 4.00, 0.00006000)
};\label{gp:IUPAitr4}

\addplot[ color=black ,mark=triangle ] coordinates {
( 2.00, 0.05113000)
( 2.25, 0.02804000)
( 2.50, 0.01444000)
( 2.75, 0.00676000)
( 3.00, 0.00284000)
( 3.25, 0.00111000)
( 3.50, 0.00045000)
( 3.75, 0.00016000)
( 4.00, 0.00007000)
};\label{gp:CPAitr4}

\addplot[ color=red ,mark=triangle ] coordinates {
( 2.00, 0.05415358)
( 2.25, 0.02888838)
( 2.50, 0.01527370)
( 2.75, 0.00694000)
( 3.00, 0.00286000)
( 3.25, 0.00111000)
( 3.50, 0.00045000)
( 3.75, 0.000158)
( 4.00, 0.0000500)
};\label{gp:CPAitr3}

\addplot[ color=blue ,mark=triangle ] coordinates {
( 2.00, 0.07502000)
( 2.25, 0.04090000)
( 2.50, 0.02131000)
( 2.75, 0.01006000)
( 3.00, 0.00408000)
( 3.25, 0.00152000)
( 3.50, 0.00061000)
( 3.75, 0.00020000)
( 4.00, 0.00007000)
};\label{gp:CPAitr2}

\addplot[ color=dgreen ,mark=triangle ] coordinates {
( 2.00, 0.36713000)
( 2.25, 0.26475000)
( 2.50, 0.17592000)
( 2.75, 0.10963000)
( 3.00, 0.06123000)
( 3.25, 0.03099000)
( 3.50, 0.01418000)
( 3.75, 0.00558000)
( 4.00, 0.00224000)
};\label{gp:CPAitr1}

		\end{groupplot}
		\node[below = 0.8cm of fer_queries c1r1.south] { (a) \footnotesize RM$(6,3)$ };
		\node[below = 0.8cm of fer_queries c2r1.south] {(b) \footnotesize RM$(7,3)$};
		\path (top|-current bounding box.north) -- coordinate(legendpos) (bot|-current bounding box.north);
		\matrix[
		matrix of nodes,
		anchor=south,
		draw,
		inner sep=0.1em,
		draw,
		column 1/.style={anchor=base west},
    	column 2/.style={anchor=base west},
    	column 3/.style={anchor=base west},
    	column 4/.style={anchor=base west},
		]at(legendpos)
		{
			\ref{gp:IUPAitr1}& \scriptsize IUPA: 1 Itr &[3pt]			
			\ref{gp:IUPAitr2}& \scriptsize IUPA: 2 Itr &[3pt]
			\ref{gp:IUPAitr3}& \scriptsize IUPA: 3 Itr &[3pt]
			\ref{gp:IUPAitr4}& \scriptsize IUPA: 4 Itr \\
			\ref{gp:CPAitr1}& \scriptsize CPA: 1 Itr &[3pt]
			\ref{gp:CPAitr2}& \scriptsize CPA: 2 Itr &[3pt]					
			\ref{gp:CPAitr3}& \scriptsize CPA: 3 Itr &[3pt]			
			\ref{gp:CPAitr4}& \scriptsize CPA: 4 Itr \\			
			};
	\end{tikzpicture}
	\caption{Comparison of different numbers of iterations for the IUPA and CPA algorithm.}
	\label{fig:iupa_itr}
\end{figure}
Since the number of iterations $N_{\max}$ for hardware implementation directly impacts resource requirements in our architecture, we explored reducing the number of iterations below $\lceil {\frac{m}{2}}\rceil$ for the CPA and IUPA decoders. 
As shown in Fig.~\ref{fig:iupa_cpa}, $N_{\max} = 1$ results in poor performance. However, for the RM$(6,3)$ code, the performance with $N_{\max} = 2$ is equivalent to that with $N_{\max}= 3$, indicating that two iterations are sufficient. 
For the RM$(7,3)$ code, $N_{\max} = 2$ results in a minor degradation of $0.05$ dB in error-correcting performance compared to three and four iterations.
Given that in our architecture the resource usage scales linearly with the number of iterations, having three iterations instead of two would increase hardware usage by approximately $50\%$, which is not justified by the marginal performance gain. Therefore, we also decided to proceed with two iterations for the IUPA and CPA decoders for RM$(7,3)$.

\subsubsection{Quantization bit-width}
\label{Sec:res:fp}
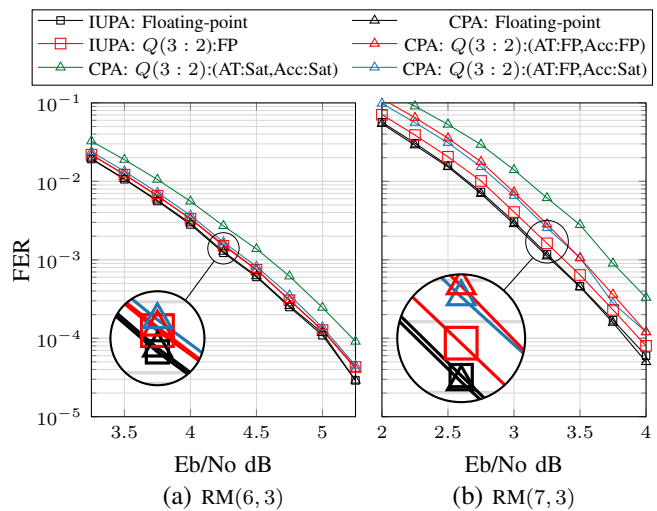
\begin{figure}[t]
	\centering
	\begin{tikzpicture}[spy using outlines={magnification=3,circle,size=1.25cm, black,connect spies}]
		\begin{groupplot}[group style={group name=fer_queries, group size= 2 by 1, horizontal sep=10pt, vertical sep=5pt},
			footnotesize,
			height=.65\columnwidth,  width=.575\columnwidth,
			xlabel=Eb\slash No dB,
			ymode=log,
			tick align=inside,
			grid=both, grid style={gray!30},
			/pgfplots/table/ignore chars={|},
			]

			\nextgroupplot[ylabel= FER, ytick pos=left, y label style={at={(axis description cs:-0.20,.5)},anchor=south},ymin=1e-5, ymax = 1e-1,xmin=3.25, xmax=5.25, xtick={2.5,3,3.5,4,4.5,5}]

\coordinate (pl1) at (axis cs: 4.25,1.4e-3);
\coordinate (pl2) at (axis cs: 3.75,1e-4);
\spy on (pl1) in node[fill=white] at (pl2);

\addplot[ color=black ,mark=square,mark options={scale=0.7} ] coordinates {
( 3.25, 0.01923400)
( 3.50, 0.01062000)
( 3.75, 0.00560700)
( 4.00, 0.00281800)
( 4.25, 0.00122500)
( 4.50, 0.00060400)
( 4.75, 0.00025000)
( 5.00, 0.00010900)
( 5.25, 0.00002900)
( 5.50, 0.00000700)
};

\addplot[ color=red ,mark=square ] coordinates {
( 3.25, 0.02188500)
( 3.50, 0.01217700)
( 3.75, 0.00653800)
( 4.00, 0.00332600)
( 4.25, 0.00150900)
( 4.50, 0.00074700)
( 4.75, 0.00030500)
( 5.00, 0.00012700)
( 5.25, 0.00004300)
( 5.50, 0.00000900)
};\label{gp:IUPA32Full}

\addplot[ color=black ,mark=triangle ] coordinates {
( 3.25, 0.01952400)
( 3.50, 0.01075800)
( 3.75, 0.00572700)
( 4.00, 0.00293000)
( 4.25, 0.00127200)
( 4.50, 0.00062100)
( 4.75, 0.00025900)
( 5.00, 0.00011700)
( 5.25, 0.00002900)
( 5.50, 0.00000800)
};

\addplot[ color=red ,mark=triangle ] coordinates {
( 3.25, 0.02183600)
( 3.50, 0.01223800)
( 3.75, 0.00654000)
( 4.00, 0.00334100)
( 4.25, 0.00156100)
( 4.50, 0.00074600)
( 4.75, 0.00030800)
( 5.00, 0.00012600)
( 5.25, 0.00004100)
( 5.50, 0.00001100)
};

\addplot[ color=Paired-1 ,mark=triangle ] coordinates {
( 3.25, 0.02385800)
( 3.50, 0.01343500)
( 3.75, 0.00719800)
( 4.00, 0.00368600)
( 4.25, 0.00167900)
( 4.50, 0.00082100)
( 4.75, 0.00035300)
( 5.00, 0.00013800)
( 5.25, 0.00004500)
( 5.50, 0.00001300)
};\label{gp:CPA32add:fp_Acc:Sat}

\addplot[ color=dgreen ,mark=triangle ] coordinates {
( 3.25, 0.03251300)
( 3.50, 0.01890700)
( 3.75, 0.01054400)
( 4.00, 0.00556300)
( 4.25, 0.00270800)
( 4.50, 0.00137700)
( 4.75, 0.00061700)
( 5.00, 0.00024600)
( 5.25, 0.00009100)
( 5.50, 0.00003000)
};\label{gp:CPA32Sat}

			\coordinate (top) at (rel axis cs:-0.2,1);

\nextgroupplot[yticklabels=\empty,ymin=1e-5, ymax = 1e-1,xmin=2, xmax=4, xtick={2,2.5,3,3.5,4}]

\coordinate (bot) at (rel axis cs:1,0);

\coordinate (p211) at (axis cs: 3.25,1.7e-3);
\coordinate (p212) at (axis cs: 2.6,1e-4);
\spy [size=1.7cm] on (p211) in node[fill=white] at (p212);


\addplot[ color=black ,mark=square,mark options={scale=0.7} ] coordinates {
( 2.00, 0.05670000)
( 2.25, 0.03069000)
( 2.50, 0.01589000)
( 2.75, 0.00733000)
( 3.00, 0.00306000)
( 3.25, 0.00118000)
( 3.50, 0.00046000)
( 3.75, 0.00017000)
( 4.00, 0.00006000)
};\label{gp:IUPAfloat}

\addplot[ color=red ,mark=square ] coordinates {
( 2.00, 0.07066000)
( 2.25, 0.03888000)
( 2.50, 0.02053000)
( 2.75, 0.01011000)
( 3.00, 0.00406000)
( 3.25, 0.00162000)
( 3.50, 0.00064000)
( 3.75, 0.00023000)
( 4.00, 0.00008000)
};

\addplot[ color=black ,mark=triangle ] coordinates {
( 2.00, 0.05415358)
( 2.25, 0.02888838)
( 2.50, 0.01527370)
( 2.75, 0.00694000)
( 3.00, 0.00286000)
( 3.25, 0.00111000)
( 3.50, 0.00045000)
( 3.75, 0.000158)
( 4.00, 0.0000500)
};\label{gp:CPAfloat}

\addplot[ color=dgreen ,mark=triangle ] coordinates {
( 2.00, 0.14552000)
( 2.25, 0.09066000)
( 2.50, 0.05307000)
( 2.75, 0.02939000)
( 3.00, 0.01400000)
( 3.25, 0.00614000)
( 3.50, 0.00279000)
( 3.75, 0.00090000)
( 4.00, 0.00033000)
};

\addplot[ color=Paired-1 ,mark=triangle ] coordinates {
( 2.00, 0.09829000)
( 2.25, 0.05606000)
( 2.50, 0.03108000)
( 2.75, 0.01529000)
( 3.00, 0.00657000)
( 3.25, 0.00255000)
( 3.50, 0.00106000)
( 3.75, 0.00028000)
( 4.00, 0.00012000)
};

\addplot[ color=red ,mark=triangle ] coordinates {
( 2.00, 0.11285000)
( 2.25, 0.06476000)
( 2.50, 0.03548000)
( 2.75, 0.01776000)
( 3.00, 0.00726000)
( 3.25, 0.00284000)
( 3.50, 0.00105000)
( 3.75, 0.00036000)
( 4.00, 0.00012000)
};\label{gp:CPA32Full}


		\end{groupplot}
		\node[below = 0.8cm of fer_queries c1r1.south] { (a) \footnotesize RM$(6,3)$};
		\node[below = 0.8cm of fer_queries c2r1.south] {(b) \footnotesize RM$(7,3)$};
		\path (top|-current bounding box.north) -- coordinate(legendpos) (bot|-current bounding box.north);
		\matrix[
		matrix of nodes,
		anchor=south,
		draw,
		inner sep=0.1em,
		draw,
		column 1/.style={anchor=base west},
    	column 2/.style={anchor=base west},
		]at(legendpos)
		{
			\ref{gp:IUPAfloat}  & \scriptsize	 IUPA: Floating-point &[3pt]
			\ref{gp:CPAfloat}& \scriptsize CPA: Floating-point \\
			\ref{gp:IUPA32Full}& \scriptsize IUPA: $Q(3:2$):FP &[3pt]	
			\ref{gp:CPA32Full}& \scriptsize CPA: $Q(3:2)$:(AT:FP,Acc:FP) \\
			\ref{gp:CPA32Sat}& \scriptsize CPA: $Q(3:2)$:(AT:Sat,Acc:Sat)  &[3pt]
			\ref{gp:CPA32add:fp_Acc:Sat}& \scriptsize CPA: $Q(3:2)$:(AT:FP,Acc:Sat) \\
			};
	\end{tikzpicture}
	\caption{FER comparison between floating-point and different fixed-point implementations of IUPA and CPA decoders.}
	\label{fig:iupa_fixed}
\end{figure}

As the final simulation step before hardware implementation, we need to determine the quantization bit width and the precision required for adders in the CPA decoder. 
Similar to the IUPA decoder~\cite{HashemipourTCAS2023}, we use 5-bit quantization for channel LLRs, with 3-bit integer and 2-bit fractional parts (represented as Q(3:2)) for IUPA and CPA decoders.
Fig.~\ref{fig:iupa_fixed} shows the performance of the IUPA decoder with 5-bit LLRs as inputs and full-precision (FP) internal additions wherever needed. Additionally, it shows the performance of the CPA decoder with Q(3:2) input LLRs and full precision for both the adder tree (AT) and accumulator (Acc). However, for CPA with $p$ PUs, having full-precision adders with $q + \log_2(p)$ bits for the adder tree and $q + \log_2\left(\binom{m}{r-1}_2\right)$ bits for the accumulator is costly. Therefore, the adder tree and accumulator defined in Section~\ref{sec:cpa:average} should be implemented to saturate to a certain value with a limited number of bits in case of overflow, instead of using full precision. The simplest option is to saturate both the adder tree and the accumulator to $q$ bits. However, this approach does not provide good error-correcting performance, as shown in Fig.~\ref{fig:iupa_fixed}. After several simulations with different quantization levels for the CPA adders, we observed that using a full-precision adder tree with $q + \log_2(p)$ bits of quantization for $p$ inputs and a saturating (Sat) accumulator yields a FER close to that of the full-precision CPA. We examined different values of $p$ (i.e., different numbers of PUs), all showing similar performance. The plots for both RM$(6,3)$ and RM$(7,3)$ codes in Fig.~\ref{fig:iupa_fixed} use $p=7$, with an 8-bit full-precision adder tree and a saturating accumulator that saturates to $\pm 2^7$ in case of overflow.
\subsection{ASIC synthesis results}
\label{sec:res:synth}
In this section, we present synthesis results for our proposed soft-input IUPA and CPA decoders, and we compare them to the IPA decoder~\cite{HashemipourTCAS2023}.
We synthesized the decoders using Cadence Genus with an STM 28nm FD-SOI technology, operating in the slow-slow corner at $25^\circ$C.
We also conducted gate-level (GL) simulations by generating a standard delay file (SDF) from synthesis to achieve precise power measurements.
This SDF was used for GL simulations in Cadence Xcelium. Moreover, we included the switching activity file obtained from the GL simulation in our analysis for power estimation.
All decoders have been implemented for 5-bit input LLRs, i.e., Q(3:2), and two decoding iterations. For the CPA decoder with $p$ PUs, the adder tree is implemented with full precision using $5 + \log_2(p)$ bits and the accumulator with saturation to $\pm 2^{4 + \log_2(p)}$. Additionally, the output of the first iteration is saturated to $\pm 2^{4}$ for the second iteration of the CPA decoder.

Table~\ref{tab:rm_6_3} and Table~\ref{tab:rm_7_3} show the synthesis results for RM$(6,3)$ and RM$(7,3)$ codes, respectively. 
We provide synthesis results for different numbers of PUs (with various $\lambda$ and $G$) of the IUPA decoder to show the trade-off between area and power consumption, latency, and throughput. 
Table~\ref{tab:rm_6_3} demonstrates that the area usage varies for the IUPA decoder when the number of groups differs, despite similar latency, throughput, and  the same number of PUs.
More specifically, with 12 PUs for the IUPA decoder for the RM$(6,3)$ code, we have two configurations for $(\lambda, G)$: $(2, 4)$ and $(4, 8)$. The ILP optimization for both ended up with 12 PUs, and they have similar throughput and latency after implementation. 
However, the area usage for the one with $G=4$ is $34\%$ higher than the one with $G=2$. As explained earlier, each second-order decoder requires its own control unit and register array, so a higher value for $G$, or a higher number of second-order decoders, results in higher resource usage. 
Therefore,  when multiple configurations  share the same location in the latency Pareto chart depicted in Fig.~\ref{fig:paretoRM73}, one should always choose the one with fewer groups.

\begin{table}[t]

\captionsetup{textfont={sc,footnotesize},justification=centerlast, labelsep=newline}
\centering
\caption{\label{tab:rm_6_3} Synthesis results for our proposed IUPA and CPA decoders compared to  SIPA for RM$(6,3)$ code.}

\begin{adjustbox}{width=\columnwidth,center}

\begin{threeparttable}
\begin{tabular}{lccccccccc}
\toprule
\textbf{Code}                 & \multicolumn{9}{c}{RM$(6,3)$}   \\ 
\textbf{Decoder}              & \multicolumn{3}{c}{IPA~\cite{HashemipourTCAS2023}}  & \multicolumn{4}{c}{IUPA} & \multicolumn{2}{c}{CPA} \\ 
\midrule
\textbf{PU}        & $16$  & $32$ & $64$ & $6$ & $12$ & $12$ & $24$ & $7$ & $21$   \\ 
\textbf{Second-Order Decoder}        & 1  & 1 & 2 & $2$  & $2$ & $4$& $2$   & - & -   \\ 
\textbf{$(G,\lambda)$}        & -  & - & - & $(2,8)$  & $(2,4)$ & $(4,8)$& $(2,2)$   & - & -   \\ 
\midrule
\textbf{Clock Rate $(\mathrm{MHz})$}    & $714$ & $714$ & $714$ & $714$ & $714$ & $714$ & $714$ & $500$ & $500$  \\ 
\textbf{Latency  $(\mathrm{cc})$ } & $294$ & $168$ & $106$ & $294$ & $164$ & $170$ & $98$ & $202$ & $78$ \\ 
\textbf{Latency $(\mathrm{\mu s})$ } & $0.411$ & $0.235$ & $0.148$ & $0.411$ & $0.229$ & $0.238$ &$0.137$ & $0.404$ &  $0.156$ \\
\textbf{Throughput $(\mathrm{Mbps})$}    & $357$ & $714$ & $1428$ & $357$ & $714$ & $714$ & $1428$  & $344$  & $1032$   \\ 
\midrule
\textbf{Area $\left(\mathrm{mm}^2\right)$}        & $0.38$ & $0.61$ & $1.21$ & $ 0.25$ & $0.32$ & $0.43$ & $0.47$ & $0.51$ &$1.21$   \\
\textbf{Area Eff. $\left(\mathrm{Gbps} / \mathrm{mm}^2\right)$} & $0.94$ & $1.17$ & $1.18 $ & $1.43$ & $2.16$ & $1.66$ & $ 3.03$ & $0.67$ & $0.85$  \\
\midrule
{\textbf{Power $(\mathrm{mW})$}} & $317$ & $505$ & $801$ & $175$ & $256$ & $316$  & $422$ & $355$ & $828 $  \\
{\textbf{Energy $(\mathrm{pJ} / \mathrm{b})$}} & $888$ & $707$ & $561$ &$490$ & $358$ & $442$ & $295$ & $1031$ & $802$  \\

\bottomrule
\end{tabular}

\end{threeparttable}

\end{adjustbox}

\end{table}
\begin{table}[t]

\captionsetup{textfont={sc,footnotesize},justification=centerlast, labelsep=newline}
\centering
\caption{\label{tab:rm_7_3} Synthesis results for our proposed IUPA and CPA decoders compared to  SIPA for RM$(7,3)$ code.}

\begin{adjustbox}{width=\columnwidth,center}

\begin{threeparttable}
\begin{tabular}{lcccccccc}
\toprule
\textbf{Code}                 & \multicolumn{6}{c}{RM$(7,3)$}   \\
\textbf{Decoder}              & \multicolumn{3}{c}{IPA~\cite{HashemipourTCAS2023}}  & \multicolumn{3}{c}{IUPA} & \multicolumn{2}{c}{CPA}\\
\midrule
\textbf{PU}        & $16$  & $32$ & $64$ & $6$ & $12$ & $24$ & $7$ & $21$   \\
\textbf{Second-Order Decoder}        & $1$  & $1$ & $1$ & $2$ & $2$ & $2$ & - & -   \\
\textbf{$(G,\lambda)$}        & -  & - & - &$(2,16)$ & $(2,8)$ &$(2,4)$& - & - \\ 

\midrule
\textbf{Clock Rate $(\mathrm{MHz})$}    & $625 $ & $625 $ & $625 $ & $625 $ & $625 $ & $625 $ & $465$ &$465$   \\
\textbf{Latency  $(\mathrm{cc})$ } & $1060 $ & $556$ & $304$ & $ 1072 $ & $ 552 $ & $292$ & $778$& $272$   \\
\textbf{Latency $(\mathrm{\mu s})$ } & $1.696 $ & $0.890 $ & $0.486 $ & $ 1.715 $ & $0.883 $ &$0.467 $ &$1.672$& $0.585$    \\
\textbf{Throughput $(\mathrm{Mbps})$}    & $157 $ & $314 $ & $628 $ & $156 $ & $312 $ & $ 625$  & $156$ & $ 469$    \\
\midrule
\textbf{Area $\left(\mathrm{mm}^2\right)$}        & $0.84$ & $ 1.48 $ & $2.60 $ & $0.62 $ & $0.79 $ & $1.18$  & $1.23$ & $2.72$  \\
\textbf{Area Eff. $\left(\mathrm{Gbps} / \mathrm{mm}^2\right)$} & $ 0.19 $ & $0.21$ & $0.24$ & $0.25 $ & $0.39 $ & $0.53$ & $0.13$ & $0.17$  \\
\midrule
{\textbf{Power $(\mathrm{mW})$}} & $ 676 $ & $1302 $ & $2221 $ & {$377$} & $562 $ & $906$ &$641$ & $1700$  \\
{\textbf{Energy $(\mathrm{pJ} / \mathrm{b})$}} & $ 4305 $ & $4146 $ & $3536$ & $2417$ &$1801$ &$1445$  & $4109$ & $3624$  \\

\bottomrule
\end{tabular}


\end{threeparttable}

\end{adjustbox}

\end{table}

Additionally, compared to the IPA decoder, approximately one-third of the PUs are needed in the IUPA for a comparable throughput and latency as shown in Table~\ref{tab:rm_6_3} and Table~\ref{tab:rm_7_3}. However, the number of second-order decoders also affects the resource usage. Therefore, the resource usage reduction in the IUPA with a higher number of second-order decoders is limited to a maximum of $54\%$  compared to the IPA with similar latency and throughput for both RM$(6,3)$ and RM$(7,3)$ codes.
However, for the configuration with the same number of second-order decoders (e.g., IUPA for RM$(6,3)$ with 24 PUs and $(G, \lambda) = (2, 2)$ and IPA with 64 PUs), the reduction in resource usage,  which is approximately $62\%$, aligns with the reduction in the number of PUs.

Moreover, we synthesized the CPA decoder with 7 and 21 PUs for both RM$(6,3)$ and RM$(7,3)$ codes to achieve similar throughput and latency with the IPA and IUPA decoders, enabling a fair comparison. 
The CPA decoding for third-order RM codes projects the received vector directly into first-order vectors using a 2-dimensional subspace. Consequently, the projection and the pre-aggregation units in the CPA are significantly larger than those in the IPA and IUPA decoders, as detailed in Section~\ref{Sec:CPA:PU}.
Following the design principle for IUPA and IPA decoder, we allocate one clock cycle for the projection unit. Due to its larger size compared to its counterpart in the IUPA and IPA decoders,  the critical path is longer and the frequency is lower. 
Allocating more clock cycles would necessitate additional registers in our pipelined architecture, leading to increased resource usage. Therefore, it is a trade-off between resource usage and frequency. Due to  the substantial number of registers required, we opted to allocate a single clock cycle for the projection unit.

The synthesis results for both RM$(6,3)$ and RM$(7,3)$ codes demonstrate that while the CPA exhibits up to a $66\%$ decrease in the number of PUs compared to the IPA for similar latency, it does not result in a substantial reduction in area. This is due to the fact that each PU in the CPA is significantly larger than in the IPA. 
Furthermore, the IPA decoder shows better overall area and energy efficiency than the CPA decoder. Specifically, the synthesis results for both RM$(6,3)$ and RM$(7,3)$ reveal that the IPA with $64$ PUs outperforms the CPA with 21 PUs in terms of latency, area, and energy efficiency.



\section{Conclusion}
\label{sec:conclusion}
In this work, we addressed the challenges associated with applying the unique projection selection method to the hardware implementation of the IPA decoder and proposed a solution based on an ILP formulation. Our ILP-based solution demonstrated a significant reduction in the number of PUs compared to the traditional IPA decoder.
We also introduced a pipelined and flexible architecture for the soft-input IUPA decoder, allowing for various configurations that balance area, power consumption, latency, and throughput.
Synthesis results demonstrated that the IUPA decoder outperforms the IPA decoder in terms of resource usage, showing a potential reduction of up to $60\%$ for certain configurations. This flexible architecture can also provide a robust foundation for further pruning and optimization of the IPA decoder.
Additionally, we implemented the soft-input CPA decoder, which has the same algorithmic complexity as the IUPA decoder to compare hardware usage. 
Synthesis results illustrate that CPA decoder is about twice 
as large as IUPA and $1.3$ times larger than IPA for similar latency and throughput.
Our findings are a reminder that algorithmic optimizations do not always translate to more efficient hardware.

%
%


\bibliographystyle{IEEEtran}
\bibliography{ref}

\end{document}